\documentstyle[prl,aps]{revtex}
\tighten
\begin{document}
\draft
%\twocolumn[
%\hsize\textwidth\columnwidth\hsize\csname @twocolumnfalse\endcsname

\title{Phase coherent transport in hybrid superconducting nanostructures.}

\author{   C.J. Lambert$^{a}$ and R. Raimondi$^{b}$}
\address{
$^{a}$School of Physics and Chemistry,
Lancaster University, Lancaster LA1 4YB, U.K.}
\address{ $^{b}$  Dipartimento di Fisica E. Amaldi, Terza Universit\`a
degli Studi di Roma, Via della Vasca Navale 84, I-00146 Roma, Italy}

\date{\today}
\maketitle

\begin{abstract}
This article is an overview of recent experimental and theoretical work
on transport in phase-coherent hybrid nanostructures, with particular emphasis
on dc electrical conduction. A summary of multiple scattering theory and
the quasi-classical methods  is presented and  comparisons
between the two are made. Several paradigms of phase-coherent transport
are discussed, including zero-bias anomalies, reentrant and long range
proximity effects, Andreev interferometers and superconductivity-induced
conductance suppression.
\end{abstract}
%]
%\narrowtext
\tableofcontents
% the introduction
\section{Introduction}
\label{introduction}

During the past few years, phase-coherent quasi-particle
transport in hybrid superconducting
structures has emerged as a new field of study, bringing together
the hitherto separate areas of superconductivity and 
mesoscopic physics [Altshuler, Lee and Webb 1991, Buot 1993].
The  seeds of this new field of research were
sown for a variety of different reasons.
The search for novel devices led several groups
to embark on experimental programmes aimed at exploring the
properties of hybrid semiconducting-superconducting structures.
During the 1980s there had been great progress in
understanding transport properties of  normal sub-micron  conductors
and therefore towards the end of that decade,
it seemed natural to ask how these were affected by the presence
of superconductivity.
Partly as a consequence of formidable technological problems
to be overcome in growing  hybrid nanostructures, the first
experimental results did not emerge until the summer of 1991, when two
groups [Petrashov and Antonov 1991, Kastalskii et al 1991]
independently submitted papers on phase-coherent transport
in sub-micron superconducting structures. The first of these reported
an {\bf increase} in resistance in metallic conductors due to the
onset of superconductivity and the second the discovery of a zero-bias
anomaly in the dc resistance of a superconductor - semiconductor (S-Sm)
contact. The former is still the subject of theoretical investigation
[Wilhelm et al 1997, Seviour et al 1997],  while the
latter is now well-understood.

In a phase-coherent
 normal-superconducting (N-S) structure,
the phase of quasi-particles as well as Cooper pairs
is preserved and
transport properties depend in detail on the quasi-particle states produced
by elastic scattering from inhomogeneities and boundaries.
A key physical phenomenon, which arises in the presence of superconductivity
is the possibility that an electron can coherently evolve into a hole and
vice versa. This phenomenon, known as Andreev scattering [Andreev 1964],
occurs without phase breaking
 and is describable by a variety of theoretical techniques.
The effect of superconductivity on transport across
a N-S interface
is of course an old subject.
In lowest order, the classical tunneling Hamiltonian
approach ignores Andreev scattering and
predicts that the dc conductance
$G$ is proportional to the density of states. Later
[Shelankov 1980, Blonder, Tinkham and
Klapwijk (BTK) 1982, Blonder and Tinkham 1983, Shelankov 1984]
it was pointed out
that the contribution to the
sub-gap conductance from Andreev scattering can be
significant and a theory of a clean N-I-S interface was developed,
which showed that for a delta-function barrier,
 there is indeed a marked deviation from
tunneling theory, but as the barrier strength is increased,
the result of classical tunneling theory is recovered.
BTK theory applies to a one-dimensional N-I-S system or, by summing
over all transverse wavevectors, to  2 or 3 dimensional systems
with translational invariance in the plane of the barrier and yields for
the current $I$ through the contact

\begin{equation}
I=(2e/h)\Omega\int^\infty_{-\infty}dE\,
(f(E-eV)-f(E))(1+A(E)-B(E))
\label{1.1}
\end{equation}
where $A(E)$ and $B(E)$ are Andreev and normal reflection coefficients
listed in table II of [Blonder et al 1982], $f(E)$ the Fermi function and
$\Omega$ a measure of the area of the junction. In the presence
of disorder or other inhomogeneities, this must be replaced by the more
general expressions outlined in sections II and IV below.

Prior to 1991, experiments on N-I-S point contacts have been in broad agreement
with BTK theory, exhibiting a conductance minimum at zero voltage $V=0$
and a peak at $eV\approx\Delta$, where $\Delta$ is the superconducting
energy gap.
In the experiment of Kastalskii et al. [1991],  the dc current through a Nb-InGaAs
contact is measured as function of the applied voltage.
At the interface, depending on the semiconductor doping level, a
Schottky barrier naturally forms so that the system
behaves like a superconductor-insulator-normal (S-I-N)
structure. (An exception to this is InAs, which does not form
a Schottky barrier at an N-S interface.) According
to BTK theory, as the barrier strength increases
the sub-gap conductance should vanish. In contrast, the experiment revealed
an excess sub-gap conductance at low bias, whose value was comparable
with the conductance arising when the superconducting
electrode is in the normal state.
This zero bias anomaly (ZBA) was later observed by Nguyen et al. [1992]
in an experiment involving InAs-AlSb
quantum wells attached to superconducting Nb contacts and
by using high transmittance Nb-Ag (or Al) contacts of varying geometry,
Xiong et al. [1993] were able to observe
the evolution from BTK to ZBA behaviour.
In an experiment by Bakker et al. [1994] involving
a silicon-based
two-dimensional electron gas (2DEG) contacted to two superconducting
electrodes, a gate voltage was also used to control
the strength of the ZBA and in [Magnee et al 1994] an extensive
 study of the ZBA in Nb/Si structures was performed.
  Since these early experiments, a great deal of effort
has been aimed at observing Andreev scattering in ballistic
2DEGs, including [van Wees et al 1994, Dimoulas et al 1995,
Marsh et al 1994, Takayanagi and Akazaki 1995(a), Takayanagi, Toyoda
and Akazaki 1996(a)]

Kastalskii et al [1991]
attributed the excess conductance to a non-equilibrium
proximity effect, in which superconductivity is induced in the
normal electrode, giving rise to an excess pair current.
Initially this phenomenon was seen as separate
from Andreev reflection,
but subsequent theoretical developments have shown that the
distinction between the proximity effect and Andreev reflection is
artificial. As will become clear later, the ZBA arises from an
interplay between Andreev scattering and disorder-induced scattering in the
normal electrode. Andreev scattering is sensitive to the breaking of
time reversal symmetry and as a consequence the
 conductance peak is destroyed by the introduction of a magnetic
field.

Zero bias anomalies constitute the first of a small number of paradigms of
phase-coherent transport in hybrid N-S structures.
A second paradigm is the observation of re-entrant
[van Wees et al 1994, Charlat et al 1996]
and long-range [Courtois et al 1996 ] behaviour
signalled by the appearance of finite-bias anomalies (FBAs)
in the conductance of high-quality N-S interfaces.
At high temperatures $T>T^*$
and bias-voltages $V>V^*$, where for
a N-metal of length $L$ and diffusion coefficient $D$,
 $k_BT^*=eV^*=\sqrt(D/L^2)$, both the ZBA and FBA conductance peaks
decay as $1/\sqrt T$ and  $1/\sqrt V$. For a clean interface there also exists a
conductance maximum at $V^*$, $T^*$ and therefore at low-temperature
and voltage a re-entrance to the low-conductance state occurs.
An interesting feature of this phenomenon is the long-range
nature of the effect, which typically decays as a power-law in $L^*/L$,
where $L^*=\sqrt(D/eV)$. This behaviour is in sharp contrast
with the exponential decay of the Josephson effect and has been observed in
a number of experiments. In a
 T-shaped Ag sample with Al islands at different distances from
the current-voltage probes [Petrashov et al 1993(b)],
a long-range proximity effect was observed, in which
the influence of the island extended over length scales greater than
the thermal coherence length $L^*$. Similar behaviour was
also observed [Petrashov et al 1994] in
ferromagetic-superconductor hybrids made from Ni-Sn and Ni-Pb.
In an experiment involving a square Cu loop in contact with 2 Al electrodes
Courtois et al [1996]
clearly identified both the short- and long-range
 contributions to phase-coherent
transport. In this interferometer experiment, they observed
a phase-periodic conductance decaying as a power-law in $1/T$,
 in parallel with a Josephson
current which decays exponentially with $L/L_T$.

The above re-entrance phenomenon is also observed in a third paradigm
of phase-coherent transport, which arises when a normal metal is in
contact with two superconductors, with order parameters phases $\phi_1$ and
$\phi_2$, whose difference $\phi=\phi_1-\phi_2$
can be varied by some external means. Prior to the experimental realisation
of these structures,
the electrical conductance
of such Andreev interferometers was predicted to be an oscillatory
function of $\phi$. Spivak and Khmel'nitskii [1982] and Al'tshuler and
Spivak [1987] identified a high temperature $(T>>T^*)$,
 weak localisation contribution to the
conductance of a disordered sample, whose amplitude of oscillation was less than or of order
$2e^2/h$. For an individual sample,
the period of oscillation was found to be $2\pi$, but for the ensemble
average a period of $\pi$ was predicted.
  Nakano and Takayanagi [1991] and Takagi [1992] examined
a clean interferometer in one-dimension and again predicted a $2\pi$-periodic
conductance with an amplitude of oscillation less than or of order
$2e^2/h$.
Lambert [1993] examined a disordered conductor in the low-temperature
limit $(T<<T^*)$ and identified a new contribution to the ensemble averaged
conductance with a periodicity of $2\pi$.
This $2\pi$ periodicity is a consequence of particle-hole
symmetry, which also guarantees that at zero temperature and
voltage, the conductance should possess a zero phase extremum [Lambert 1994].
Prior to experiments on such devices,
the generic nature of this prediction
was confirmed in numerical simulations [Hui and Lambert 1993(a)] encompassing
the ballistic, diffusive and almost localised regimes and in a
tunnelling calculation of the ensemble averaged conductance
by Hekking and Nazarov [1993]. 

The first experimental realisations of Andreev interferometers
came  almost simultaneously from three separate groups.
In March of 1994, de Vegvar et al [1994] showed results
for a structure formed from two Nb electrodes
in contact with an Al wire. They found a small
oscillation $10^{-3}(2e^2/h)$ with a sample specific phase in the
$2\pi$ periodic component, suggesting that the ensemble
averaged conductance should have a periodicity of
$\pi$, in agreement with Spivak and Khmel'nitskii [1982].
However in contrast with all subsequent experiments,
no zero phase extremum was observed.
In April/May of that year, Pothier et al [1994] produced an interferometer
involving two tunnel junctions, which showed a $2\pi$-periodic conductance,
with a zero phase maximum and a low-bias, low-temperature amplitude
of oscillation of order $10^{-2}(2e^2/h)$, which decayed with increasing
temperature. 
In May 1994, van Wees et al [1994], [see also Dimoulas et al 1995]
produced the first quasi-ballistic InAs 2DEG interferometer, with high
transparency N-S interfaces. This experiment showed the first
re-entrant behaviour in which the amplitude of oscillation
$\delta G$ varied from $\delta G\approx -0.08 (2e^2/h)$  at zero
voltage,( where a minus sign indicates a zero phase minimum
and a + sign a zero-phase maximum) passes through zero at a bias of
order 0.1mV, reaches a maximum at a bias of order $V^*$
 and then decays to zero at higher voltages. (For a detailed study
 see [den Hartog et al, 1996]).
Unlike the Josephson current which decays exponentially with $T/T^*$,
these conductance oscillations decayed only as a power-law.

The first experiment showing an amplitude of oscillation
greater than $2e^2/h$ was carried out by
Petrashov et al [1995]. Here, silver or antimony wires
in the shape of a cross, make two separate contacts with 
superconducting Al and the phase difference between the contacts
is varied using either
an external field applied to a superconducting loop or by passing a
supercurrent throught the Al. The amplitude was found to be
$\delta G \approx 100(2e^2/h)$ for Ag
and $3.10^{-2}(2e^2/h)$ for Sb, and exhibited a periodicity of $2\pi$.
In this experiment, the phase difference $\phi$ was varied both by
passing a magnetic flux through an external superconducitng loop and by
passing a supercurrent through a straight section of the superconductor,
thereby emphasising that precise the manner in which the
order parameter phase is controlled is not important.
These experiments were crucial in demonstrating that in metallic samples,
the ensemble averaged conductance is the relevant quantity and therefore
a quasi-classical description is relevant.
It is perhaps worth mentioning that with hindsight,
an earlier experiment
reporting large-scale oscillations in a sample with two
superconducting islands [Petrashov et al 1993(a)]  can be regarded
a precursor to these interferometer experiments. However the phase
of the islands was not explicitly controlled, making an interpretation of
the 1993 experiments more difficult.

A fourth and more recent paradigm of phase-coherent transport
is the appearance of negative multi-probe conductances in
structures where Andreev transmission of quasi-particles is a dominant
process [Allsopp et al 1994]. The first experiments reporting this behaviour
were carried out by Hartog et al [1996], using a diffusive InAs 2DEG.
These probe individual coefficients in the current-voltage
relations and  demonstrate
fundamental reciprocity relations arising from time-reversal and
particle-hole symmetry.

Finally a fifth
paradigm is the suppression of electrical
 conductance by superconductivity in metallic systems without tunnel barriers.
  Experimentally this phenomenon
 has been observed in several structures, involving both non-magnetic
 (ie Silver) and magnetic (ie Nickel) N-components in contact a
 superconductor via clean interfaces.
In the experiment of Petrashov and Antonov [1991]
the conductance of a Ag wire on which several Pb islands are deposited,
decreases by 3\% when the islands become superconducting,
this corresponding to a conductance decrease of
$\delta G\approx 100~(2e^2/h)$.
These experiments reveal two regimes: as the field decreases below ${\rm Hc}_2$,
an initial
resistance increase occurs. The resistance remains field insensistive until
a lower field is reached (corresponding to the order of
a flux quantum through the sample),
at which point a further conductance suppression occurs.
In the experiment by Petrashov et al [1993b] involving a
 T-shaped Ag sample with Al islands at different distances from
the current-voltage probes, a superconductivity induced change in the
resistance by up to 30\% was observed, but it was found that
for different samples, the resistance could either increase or decrease.
Although superconductivity-induced conductance suppression
in metallic samples was predicted a number of years ago [Hui and Lambert, 1993(b)]
these experiments at first sight appeared to conflict with
quasi-classical theories, which universally
 predict that the normal-state,
  zero-temperature, zero-bias conductance $G_N$ is identical
 the conductance $G_{NS}$ in the superconducting state.
This effect is addressed in [Hui and Lambert 1993b, Claughton et al 1995,
 Wilhelm et al 1997, Seviour et al 1997].

The main aim of this review is to outline the quasi-classical and multiple
scattering theories needed to describe the dc electrical conductance
of phase-coherent hybrid N-S structures. For this reason we shall not discuss
thermodynamic phenomena such as the Josephson effect in any detail,
despite the fact that ground-breaking experiments using clean
[Takayanagi and Akazaki, 1995(b,c,d)]
superconducting quantum  S-2DEG-S point contacts
show quantization of the critical current
as predicted by  Furasaki et al [1991] and
for shorter junctions by Beenaker and van Houten [1991].
There are several notable theoretical papers addressing Andreev scattering
in such structures, including [van Wees et al 1991, Bagwell 1992,
Furasaki et al 1994, Gusenheimer and Zaikin 1994,
Zyuzin 1994,
Hurd and Wendin 1994 and 1995, Bratus et al 1995, Chang et al 1995,
Koyama, Takane and Ebisawa 1995 and 1996,
Levy Yeyati et al 1996, Martin-Rodero 1996, Wendin and Shumeiko 1996,
Reidel et al 1996, Volkov and Takayanagi 1996(b)].
Similarly to restrict the length of this review,
recent theories  of thermoelectric coefficients
 [ Bagwell and Alam 1992, Claughton and Lambert 1996] and
 shot noise will not be discussed [Datta 1995],
 nor will we discuss work on coulomb effects in superconducting islands
 [Eiles et al 1993, Lafarge et al 1993, Tuominen et al 1992 and 1993,
 Hergenrother et al 1994, Black et al 1996, Hekking et al 1993].

While Andreev interference effects are generic phenomena, their manifestation
in a given experiment is sensitive to many parameters.
For the purpose of writing this review, it is therefore convenient to adopt
a simple classification of experimental
arrangements sketched in figure 1.
The generic structure shown in figure 1a represents our first class of
N-S-N hybrids and has many realisations.
For convenience, we distinguish these from a second class
of N-S hybrids of the kind
shown in figure 1b and 1c, in which the superconductor effectively
forms part of an
external reservoir and is not simply part of the scattering region.
Figure 1d indicates a third class of N-SS'-N structures,
 involving two (or more)
separate superconductors S and S$^\prime$,
with respective order parameter phases $\phi$, $\phi'$.
As noted above, transport properties of such Andreev interferometers
are periodic functions of the
phase difference $\phi-\phi'$.
Figure 1e shows a fourth class of S-N-S' structures, in which
two superconducting reservoirs are connected to a normal scattering region.
In this case, the structure forms a Josephson junction and
in contrast with all other structures shown in figure 1,
in the linear response limit, the dc conductance measured between
the superconducting reservoirs is identically zero. In this case, the
relevant dc quantity is the current-phase relation and the associated
critical current.
Clearly one could also measure the Josephson current between the
two superconductors in the N-SS'-N structure of figure 1c and therefore
the above classification is intended to label the measurement being
made, rather than the device being measured. Structures of the form 1e
will not be discussed.

For the most part, the theoretical descriptions discussed below
have finessed problems
of self-consistency, by computing measured quantities as a function of
the superconducting
 order parameter pairing field
$f_{\sigma\sigma'}(\underline r)=
\langle \psi_\sigma(\underline r)\psi_{\sigma^\prime}(\underline r)\rangle$
induced by making contact with a piece of
superconductor.
As an example, figure 1a shows a normal
mesoscopic scattering region in  contact with a superconducting island
which plays the role of an externally controllable
 source of $f_{\sigma\sigma'}(\underline r)$,
in much the same way that the coils of a magnet are an external source
of magnetic field. The coils are not of primary interest
and in many cases, neither is the superconductor.
It is assumed that parameters characterizing the
superconductor are given and the key question is how does superconductivity
influence transport through the scattering region.
Of course once the influence of superconductivity is understood,
transport properties can be used to probe the symmetry and spatial
stucture of the order parameter, as suggested by Cook et al. [1996].
 Furthermore, in the presence
of large currents which modify the order parameter, a complete
self-consistent treatment is necessary, as described for example
in [Bruder 1990, Hara et al 1993,
Barash et al 1995,
 Caniz\~ares and Sols 1995, Chang et al 1995, Gyorffy 1995,
Martin and Lambert 1995 and 1996, Riedel et al 1996].

% current-voltage relations and the bdg equation

\section {The multiple scattering approach to dc transport in superconducting hybrids.}
\subsection{Fundamental current-voltage relations.}
In this section we review the multi-channel
current-voltage relations for a disordered
phase-coherent scatterer connected
to normal reservoirs, obtained for two normal probes by [Lambert 1991]
and extended to multi-probes by
[Lambert Hui and Robinson, 1993]. The coefficients (denoted $A_{ij}$
 and $a_{ij}$
below) are directly accessible experimentally and can be combined to
yield a variety of transport coefficients, the simplest of which is
the electrical conductance of a normal-superconducting interface.
To avoid time-dependent order parameter
phases varying at the Josephson frequency, which would render
a time-independent scattering approach invalid, these are derived
under the condition that all
superconductors share a common condensate chemical potential $\mu$.
The derivation of the fundamental current-voltage relation 
presented in [Lambert 1991] follows closely the
multi-channel scattering theory developed during the 1980s for
non-superconducting
mesoscopic structures [B\"uttiker 1986, Buot 1993].
 In the normal state, this approach
yields, for example, the multi-channel
Landauer formula [Landauer 1970] for the electrical conductance

\begin{equation}
G=(2e^2/h) T_0 ,
\label{a1} 
\end{equation}

where $T_0$ is the transmission coefficient of the structure.
Historically the above formula was not accepted without
a great deal of debate and as we shall see in the following section,
contradicts the corresponding expression used by practitioners of
quasi-classical theories, where the alternative expression 

\begin{equation}
G=(2e^2/h) (T_0/R_0)
\label{a2}
\end{equation}

is employed, with (in one-dimension) $R_0$ the reflection coefficient.
In fact the above two expressions  refer to a two-probe and to a four-probe measurements, respectively,
and the crucial lesson from the debate surrounding these equations
is that transport coefficients such as the electrical conductance are
secondary quantities. More fundamental are the current-voltage
relations describing a given mesoscopic structure.

Equations (\ref{a1}) and (\ref{a2})
 are not valid in the presence of Andreev scattering,
because charge transport and quasi-particle diffusion are no longer
equivalent. For example when a quasi-particle Andreev reflects
at an N-S interface, the energy and probability density of the excitation
are reflected back into the normal conductor, whereas a charge of $2e$ is
injected into the superconductor. Thus charge flows into the
superconductor, even though the excitation does not and as a consequence,
a current-voltage relation  should be used, which takes into account
this charge-energy separation.
For a scattering region connected to $L$ normal reservoirs, labelled
$i = 1,2, \dots L$, it is convenient to write this in the form

\begin{equation}
I_i=\sum_{j=1}^{N}  A_{ij}, 
\label{a3}
\end{equation}

where $I_i$ is the current flowing from reservoir $i$ and the coefficients
$A_{ij}$
will be discussed in detail below.
In the linear-response limit, this reduces to

\begin{equation} 
I_i=\sum_{j=1}^La_{ij}( v_j- v),
\label{a4}
\end{equation}

The above expressions describe
reservoirs at voltages $v_i, i= 1,2, \dots, L$,
connected to a scattering region containing one or more superconductors
with a common condensate chemical $\mu$
and relates the current $ I_i$  from reservoir $i$
to the voltage differences  $(v_j -v)$, where $v=\mu/e$.
The $L=2$ formula
 describes a wide variety of experimental
measurements and underpins many subsequent theoretical descriptions
of disordered N-S interfaces and inhomogeneous structures.
For this reason, after discussing the relationship
between the coefficients $A_{ij}$, $a_{ij}$ and the scattering matrix,
we shall examine the  two-probe formula
in some detail and illustrate its application
to some generic experimental measurements.

\subsection{Relationship between the
generalised conductance matrix and the s-matrix.}
The $L=2$ analysis of [Lambert 1991] is based on
the observation that in the absence of inelastic scattering,
dc transport is determined by the
quantum mechanical scattering matrix $s(E,H)$, which yields scattering
properties at energy $E$, of
 a phase-coherent structure
described by a Hamiltonian $H$.
If the structure is connected to external reservoirs by open
scattering channels labelled by quantum numbers $n$, then this has matrix
elements of the form $s_{n,n'}(E,H)$. The squared modulus of $s_{n,n'}(E,H)$
is the outgoing
flux of quasi-particles along channel $n$, arising from a unit incident
flux along channel $n'$.
Adopting the notation of [Lambert, Hui and Robinson 1993],
we consider
channels belonging to current-carrying leads, with quasi-particles
labelled by a discrete quantum number $\alpha$
($\alpha = + 1$ for particles, $-1$ for holes)
and therefore
write $n=(l,\alpha)$, where $l$ labels all other quantum numbers
associated with the leads. With this notation, the
scattering matrix elements $ s_{n,n'}(E,H)=
s_{l,l'}^{\alpha,\beta}(E,H)$
satisfy the unitarity condition
 $s^\dagger(E,H)=s^{-1}(E,H)$, the time-reversibility condition
$s^t(E,H)=s(E,H^*)$ and if $E$ is measured relative to the
condensate chemical potential $\mu=ev$, the particle-hole symmetry relation
$s_{l,l'}^{\alpha,\beta}(E,H)=\alpha\beta 
[s_{l,l'}^{-\alpha,-\beta}(-E,H)]^*$.
(For convenience we adopt the convention of including appropriate
ratios of channel group velocities in the definition of $s$ to
yield a unitary scattering matrix.)

For a scatterer connected to external reservoirs by $L$ crystalline, normal
leads, labelled $i = 1,2, \dots , L$, it is
convenient to write $l=(i,a)$,
$l'=(j,b)$, where $a (b)$ is a channel belonging to lead $i (j)$.
With this notation, the quantities entering the current-voltage relation are
of the form
\begin{equation} P^{\alpha,\beta}_{i,j}(E,H)=\sum_{a,b}\vert
s_{(i,a),(j,b)}^{\alpha,\beta}(E,H)\vert^2
={\rm Tr}\,\, [s_{ij}^{\alpha,\beta}(E,H)
\{s_{ij}^{\alpha,\beta}(E,H)\}^\dagger]
, \label{a5}\end{equation}
which is an expression for the coefficient for reflection ($i=j$) or
transmission
($i\ne j$) of a quasi-particle of type $\beta$ in lead $j$ to a
quasi-particle of type $\alpha$ in lead $i$. For $\alpha \ne\beta$,
$ P^{\alpha,\beta}_{i,j}(E,H)$ is
an Andreev scattering coefficient, while for
$\alpha=\beta$, it is a normal scattering coefficient.
Since unitarity yields
\begin{equation}\sum_{\beta b j}\vert s_{(i,a),(j,b)}^{\alpha,\beta}(E,H)
\vert^2=\sum_{\alpha a j}\vert s_{(i,a),(j,b)}^{\alpha,\beta}(E,H)
\vert^2=1\label{a6},\end{equation}
where $i$ and $j$ sum  over all leads containing open channels of energy $E$,
this satisfies
\begin{equation}\sum_{\beta  j}^LP_{ij}^{\alpha,\beta}(E,H)
=N_i^\alpha(E),\,\,\, {\rm and}\,\,\,\,\,
\sum_{\alpha i}^LP_{ij}^{\alpha,\beta}(E,H)
=N_j^\beta(E)\label{a7},\end{equation}
where  $N^\alpha_i(E)$ is the number
of open channels for $\alpha$-type quasi-particles of energy $E$ in lead $i$,
satisfying $N^+_i(E)=N^-_i(-E)$. Similarly particle-hole symmetry yields
\begin{equation}P_{ij}^{\alpha,\beta}(E,H)=P_{ij}^{-\alpha,-\beta}(-E,H)\label{a8}\end{equation}
and time reversal symmetry
\begin{equation}P_{ij}^{\alpha,\beta}(E,H)=P_{ji}^{\beta,\alpha}(E,H^*)\label{a9}.\end{equation}

Having introduced the scattering coefficients 
$P_{ij}^{\alpha,\beta}(E,H)$,
the coefficients $A_{ij}$ of the fundamental formula
 (\ref{a3}) are given by

\begin{equation} A_{ij}=(2e/h)\sum_{\alpha}(\alpha)\int^\infty_0
\,dE\,\{\delta_{ij}N_i^\alpha(E)f^\alpha_i(E)
-\sum_\beta P_{ij}^{\alpha\beta}(E,H)
f^\beta_j(E)\}\label{a27},\end{equation}
with $f^\alpha_j(E)=
\{ {\rm exp}[(E-\alpha(ev_j-\mu))/k_bT]+1\}^{-1}$
the distribution of
incoming $\alpha$-type quasi-particles from lead $j$.
\footnote{
It is perhaps worth noting that in [Lambert 1991], the following notation
is employed
$N_p(E)=N^+_1$, $\tilde R_{pp}=P^{++}_{11}(E,H)/N_p(E)$,
$\tilde R_{hp}=P^{-+}_{11}(E,H)/N_p(E)$ etc.}

Equation (\ref{a3}) yields the current-voltage characteristics of a given
structure at finite voltages, provided all scattering coefficients
are computed in the presence of a self-consistently determined order
parameter and self-consistent values of all other scattering
potentials.
At finite temperature, but zero voltage,
 where $v_i-v \rightarrow 0$, it reduces
to equation (\ref{a4}), with
$a_{ij}$  given by
\begin{equation} a_{ij}=(2e^2/h)\int^\infty_{-\infty}
\,dE\,[-{\partial f(E)
\over\partial E}][N_i^+(E)\delta_{ij}-P_{ij}^{++}(E,H)
+P_{ij}^{-+}(E,H)]\label{a10},\end{equation}
where $f(E)$ is the Fermi function and equation (\ref{a8}) has been used.
At finite voltages,
but zero-temperature, it reduces to
\begin{equation}A_{ij}=(2e/h)\int^{(ev_i-\mu)}_0
\,dE\,[\delta_{ij}N_i^+(E)+P_{ij}^{-+}(E,H)-P_{ij}^{++}(E,H)]\label{a29}.\end{equation}
Finally at finite voltages,
the differential of equation (\ref{a3}) with respect to
$v_j$ (with $\mu$ and all other potentials
held constant)
yields
\begin{equation}\partial I_i/\partial v_j=  a_{ij} \label{a30},\end{equation}
where at finite temperature,

\begin{equation}
 a_{ij}=(2e^2/h)\sum_{\alpha}(\alpha)\int^\infty_0
\,dE\,\{\delta_{ij}N_i^\alpha(E)[-\alpha{\partial f_i^\alpha(E)\over\partial E}]
-\sum_\beta P_{ij}^{\alpha\beta}(E,H)
[-\beta{\partial f_j^\beta(E)\over\partial E}]\}
\label{a31}.
\end{equation}
and at zero temperature,
\begin{equation} a_{ij}=
[2e^2/h][\delta_{ij}N_i^+(E_i)+P_{ij}^{-+}(E_j,H)-P_{ij}^{++}(E_j,H)]
\label{a32}.\end{equation}
where  where $E_i=ev_i-\mu$.

It is worth noting that
replacing $E$ by $-E$ and
utilizing the particle-hole symmetry relation (\ref{a8}) allows
equation (\ref{a10}) to be rewritten in the form
\begin{equation} a_{ij}=(2e^2/h)\int^\infty_{-\infty}
\,dE\,[-{\partial f(E)
\over\partial E}][N_i^-(E)\delta_{ij}-P_{ij}^{--}(E,H)
+P_{ij}^{+-}(E,H)]\label{a19},\end{equation}
which demonstrates that particles and hole are treated on an equal footing
in equations (\ref{a10}) and (\ref{a19}).
Furthermore in view of the symmetries (\ref{a8}), (\ref{a9})
the reciprocity relation
$a_{ij}(H)=a_{ji}(H^*)$ is satisfied.

\subsection{Two probe formulae in more detail.}
While the above notation is convenient for arbitrary $L$, it perhaps
obscures the simplicity of the final result and therefore in the literature,
several alternative notations have been employed.
For the case of $L=2$ normal probes, where the scattering matrix
 has the structure
\begin{equation}
S(E,H) =
\left(\matrix{r(E) & t^\prime(E) \cr
t(E) & r^\prime(E) }\right) , 
\label{a110}
\end{equation}
it is convenient to write
$s^{\alpha\beta}_{11}(E,H)=r_{\alpha\beta}(E)$,
$s^{\alpha\beta}_{22}(E,H)=r^\prime_{\alpha\beta}(E)$,
$s^{\alpha\beta}_{21}(E,H)=t_{\alpha\beta}(E)$ and
$s^{\alpha\beta}_{12}(E,H)=t^\prime_{\alpha\beta}(E)$.
With this notation, the sub-matrices $r,t,r',t'$ have the form

\begin{equation}
r(E)  =
\left(\matrix{r_{++}(E) & r_{-+}(E) \cr
r_{-+}(E) & r_{--}(E) }\right), \,\,\,\,\,\,\,\,\,{\rm etc.}
\label{a11},\end{equation}
The matrix $r_{\alpha\beta}$ ($r^\prime_{\alpha\beta}$) is a matrix of
 amplitudes describing the reflection of  $\beta$-type quasi-particles
 from reservoir 1 (2)  into $\alpha$-type quasi-particles
travelling back into reservoir 1 (2). Similarly
$t_{\alpha\beta}$ ($t^\prime_{\alpha\beta}$) is a matrix of
 amplitudes describing the transmission of  $\beta$-type quasi-particles
 from reservoir 1 (2)  into $\alpha$-type quasi-particles
of reservoir 2 (1).

Using these sub-matrices, the $L=2$  current-voltage relation
can be written
\begin{equation}\left(\matrix{I_1\cr  I_2}\right)=
\left(\matrix{a_{11}&a_{12}\cr a_{21}&a_{22}}\right)
\left(\matrix{v_1- v\cr  v_2- v}\right)\label{a15},\end{equation}
where

\begin{equation}\left(\matrix{a_{11}&a_{12}\cr a_{21}&a_{22}}\right)=
\int^\infty_{-\infty}\, dE\,(-{{\partial f(E)}\over{\partial E}})\left(\matrix{N^+_1(E)- R_0(E)+R_a(E)
&T^\prime_a(E)-T^\prime_0(E)\cr T_a(E)-T_0(E)&N^+_2(E)-R^\prime_0(E)+R^\prime_a(E)}\right)
\label{a16},\end{equation}
with
\begin{equation}\left(\matrix{R_0(E)\cr T_0(E)\cr R_a(E)\cr T_a(E)}\right)=
\left(\matrix{P^{++}_{11}(E,H)\cr P^{++}_{21}(E,H)\cr
P^{-+}_{11}(E,H)\cr P^{-+}_{21}(E,H)}\right)=
\left(\matrix{{\rm Trace}\,\,\,\{ r_{++}(E)r^\dagger_{++}(E)\}\cr
{\rm Trace}\,\,\, \{t_{++}(E)t^\dagger_{++}(E)\}\cr
{\rm Trace}\,\,\, \{r_{-+}(E)r^\dagger_{-+}(E)\}\cr
{\rm Trace}\,\,\, \{t_{-+}(E)t^\dagger_{-+}(E)\}}\right)
\label{a17}\end{equation}
and
\begin{equation}\left(\matrix{R^\prime_0(E)\cr T^\prime_0(E)\cr R^\prime_a(E)\cr T^\prime_a(E)}\right)=
\left(\matrix{P^{++}_{22}(E,H)\cr P^{++}_{12}(E,H)\cr
P^{-+}_{22}(E,H)\cr P^{-+}_{12}(E,H)}\right)=
\left(\matrix{{\rm Trace}\,\,\, \{r^\prime_{++}(E)r^{\prime^\dagger}_{++}(E)\}\cr
{\rm Trace}\,\,\, \{t^\prime_{++}(E)t^{\prime^\dagger}_{++}(E)\}\cr
{\rm Trace}\,\,\, \{r^\prime_{-+}(E)r^{\prime^\dagger}_{-+}(E)\}\cr
{\rm Trace}\,\,\, \{t^\prime_{-+}(E)t^{\prime^\dagger}_{-+}(E)\}}\right)
\label{a18}.\end{equation}
Similarly the zero temperature differential conductance (\ref{a32}) becomes
\begin{equation} a_{ij}=
[2e^2/h][\delta_{ij}N_i^+(E_i)+R_a(E_j)-R_0(E_j)]
\label{a32a}.\end{equation}

\subsection{Applications of the two-probe conductance matrix.}
As noted in [Lambert 1991], the two-probe
current-voltage relation (\ref{a15}) can
be used to derive generalisations of both (\ref{a1}) and (\ref{a2}). To
further illustrate the
versatility of the two-probe theory, we now apply it
to some typical experimental measurements. For convenience in
what follows, we set $2e^2/h$
equal to unity.
\vskip 0.3cm
\noindent
{\bf Example 1. The normal limit.}

Consider the structure of figure 1a, where a scattering region containing
superconducting inclusions is connected to normal reservoirs at potentials
$v_1$, $v_2$. In the normal limit, the condensate potential $v$ must
disappear from the fundamental current-voltage relation (\ref{a3}). As shown below,
this occurs, because in the absence of Andreev scattering, unitarity
of the scattering matrix implies that $a_{11}=-a_{12}=-a_{21}=a_{22}=T_0$.
Hence equation (\ref{a15}) reduces to
\begin{equation} I_1=-I_2=T_0(v_1- v_2)\label{a20},\end{equation}
which is simply the Landauer formula (\ref{a1}).

\vskip 0.3cm
\noindent
{\bf Example 2. Experiments where ${\bf \mu_2 = \mu}$.}

Figure 1b shows an experiment in which the superconductor
and reservoir 2 are held at the same potential. In this case, equation (\ref{a15})
yields
\begin{equation}{{ I_1}\over{(v_1- v)}} =a_{11}
=\int^\infty_{-\infty}\, dE\,
-{{\partial f(E)}\over{\partial E}}[N^+_1(E)- R_0(E)+R_a(E)]
\label{a21}.\end{equation}
This experimental configuration is of the type used in
tunneling experiments, aimed at probing  the proximity
effect in the vicinity of an N-S boundary [Gueron et al 1996].
This result describes any of the structures shown in figure 1,
provided $\mu_2=\mu$ and
 is a generalization to disordered and inhomogeneous
structures of the boundary conductance formula derived by
[ Blonder, Tinkham, Klapwijk 1982].

\vskip 0.3cm
\noindent
{\bf Example 3. Experiments where ${\bf a_{12}=a_{21}=0}$.}

An example of such an experiment is shown in figure 1c, where under
sub-gap conditions, the presence of a long superconductor (of length
greater than the superconducting coherence length) prevents the
transmission of quasi-particles from reservoir 1 to reservoir 2 and vice versa.
In this case, combining the unitarity condition
$N^+_1(E)= R_0(E)+R_a(E)$ with equation (\ref{a15}) yields
\begin{equation}{{ I_1}\over{(v_1- v)}} =a_{11}
=\int^\infty_{-\infty}\, dE\,
\left(-{{\partial f(E)}\over{\partial E}}\right)[2 R_a(E)]
\label{a22}.\end{equation}

In common with the current-voltage relation from which it derives,
equation (\ref{a22}) is valid
 in the presence of disorder and inhomogeneities and in the presence
 of an arbitrary number  of superconducting inclusions of arbitrary geometry.

\vskip 0.3cm
\noindent
{\bf Example 4. Experiments where ${\bf I_1=-I_2=I}$.}

Such a situation is illustrated in figure 1a.
In this case, inverting equation (\ref{a15}) yields
\begin{equation}\left(\matrix{  v_1-  v\cr   v_2-  v}\right)=
{1\over d}\left(\matrix{a_{22}&-a_{12}\cr-a_{21}&a_{11}}\right)
\left(\matrix{  I\cr -  I}\right)\label{a23},\end{equation}
where
$d=a_{11}a_{22}-a_{12}a_{21}$.
Hence the two-probe conductance
$G=  I/(  v_1-  v_2)$ takes the form
\begin{equation}
G={d\over a_{11}+a_{22}+a_{12}+a_{21}}\label{a24}.\end{equation}

As an example of this formula, we note that in
the zero-temperature limit, where
all quantities are evaluated at zero energy, equation (\ref{a24})
 can be written [Lambert 1993, Lambert, Hui and Robinson 1993]
\begin{equation}G=T_0+T_a + {{2(R_a R'_a -T_a T'_a)}\over {R_a+R'_a+T_a+T'_a}}
\label{a25}.\end{equation}
For a symmetric scatterer, where  primed quantities equal unprimed quantities,
this reduces to
$G=T_0+R_a$,
whereas in the absence of transmission between the reservoirs,
the resistance $G^{-1}$ reduces to a sum of two resistances
$G^{-1}=(1/2R_a) + (1/2R^\prime_a)$. It should be noted that
a combination of particle-hole symmetry  and unitarity
yield at $E=0$, $T_0+T_a=T^\prime_0+T^\prime_a$ and therefore equation (\ref{a25})
is symmetric under an interchange of primed and unprimed coefficients.

\vskip 0.3cm
\noindent
{\bf Example 5. Experiments where ${\bf I_2=0}$.}

As a final example, consider the experiment sketched in figure 1d
where reservoir 2 acts as a voltage probe, with $I_2=0$.
In this case equation (\ref{a15}) yields for the ratio of the voltages
\begin{equation}{{ (v_2-v)}\over{ (v_1-v)}}=-{{a_{21}}\over{a_{22}}}
\label{a26},\end{equation}
where from equation (\ref{a16}), the coefficient $a_{22}$ is positive. In contrast,
the coefficient
$a_{21}$ is necessarily negative for a normal system,
but in the presence of Andreev scattering can have arbitrary sign.
Hence superconductivity can induce voltage sign-reversals which are
not present in the normal limit. This feature was first predicted
within the context of
negative four-probe conductances
[Allsopp et al 1994] and has been confirmed in recent
experiments by the Groningen group [Hartog et al 1996].

\subsection{The Bogoliubov - de Gennes equation.}

The above formulae relate measurable quantities to scattering
matrix elements and therefore
to end this section we briefly introduce the Bogoliubov - de Gennes
 equation [de Gennes 1989],
 which forms a basis for computing the scattering
 matrix $s$.
The Bogoliubov - de Gennes equation arises during the
diagonalization of the mean-field BCS Hamiltonian,
 which for a non-magnetic, spin-singlet superconductor takes the form
\begin{equation}{H}_{eff} = E_0+\int d \underline r
\int d \underline r'
\left(\psi_{\uparrow}^{\dagger}(\underline r)
\psi_{\downarrow} (\underline r)\right)
H(\underline r,\underline r')\left(\matrix{\psi_{\uparrow}
(\underline r') \cr
\psi_{\downarrow}^{\dagger} (\underline r')} \right)\label{a135},\end{equation}
\noindent
where $E_0$ is a constant,
$\psi_\sigma(\underline r)$ and $\psi^\dagger_\sigma
(\underline r)$ are field operators, destroying and creating electrons
of spin $\sigma$ at position  $\underline r$
and
\begin{equation}H(\underline r,\underline r') = \left(\matrix{\delta(
\underline r - \underline r') {H_0}(\underline r)
& \Delta (\underline r, \underline r') \cr
\Delta^*(\underline r, \underline r')
& -\delta(
\underline r - \underline r'){H_0}^*(\underline r )}\right).\label{a36}
\end{equation}
To each positive eigenvalue $E_n$ of $H$, with eigenvector
$\underline\Psi_n(\underline r)=\left( \matrix{u_n
(\underline r) \cr v_n(\underline r)}\right)$
satisfying

\begin{equation}\int d \underline r' H(\underline r, \underline r')
\left( \matrix{ u_n (\underline r') \cr
v_n (\underline r') } \right ) = E_n \left( \matrix{u_n
(\underline r) \cr
v_n (\underline r) } \right ) \label{a38},\end{equation}
 there exists a
corresponding negative eigenvalue $- E_n$ with
eigenvector
$\Psi_{-n}(\underline r)=\left( \matrix{-v^*_n
(\underline r) \cr
u^*_n (\underline r) } \right )$.
Consequently $\hat H_{eff}$ is diagonalized by the transformation
\begin{equation}\left(\matrix{\psi_\uparrow(\underline r)\cr\psi^\dagger_\downarrow(
\underline r)}\right)=\sum_{n>0}\left(\matrix{u_n(\underline r)&
-v^*_n(\underline r)\cr
v_n(\underline r)& u^*_n(\underline r)}\right)
\left(\matrix{\gamma_{n\uparrow}\cr\gamma^\dagger_{n\downarrow}
}\right)\label{a40},\end{equation}
where to avoid overcounting, only one of
 $\underline\Psi_n(\underline r)$ or $\underline\Psi_{-n}(\underline r)$ is
 included in the sum over $n$.

It should be noted that whereas the Bogoliubov - de Gennes equation
(\ref{a38}) refers to a
model in which the variable $\underline r$ varies continuously,
in a tight-binding model, often used in numerical simulations,
the corresponding 
Bogoliubov - de Gennes equation is

\begin{equation}
\begin{array}{c c}
E\psi_i
=&\epsilon_i \psi_{i}
-\gamma\sum_{\delta}    \psi_{i+\delta}
+  \Delta_{i} \phi_{i}\\ 
E\phi_i =&-
\epsilon_i \phi_{i}+\gamma^*\sum_{\delta}    \phi_{i+\delta}
+ \Delta^*_{i}\psi_{i} 
\end{array}
\label{a41}
\end{equation}

where $\psi_i$ ($\phi_i$) indicates the particle (hole) wavefunction
on site $i$ and $i+\delta$ labels a neighbour of $i$.

As discussed in [Hui and Lambert 1990] and
[Lambert, Hui and Robinson 1993],
for a scattering region connected to two crystalline normal leads,
the
Bogoliubov-de Gennes equation may be solved by means of  a
transfer matrix method, which
yields a transfer matrix $T$ satisfying

\begin{equation}\left(\matrix {O^\prime \cr I^\prime} \right) = T\left(\matrix {I \cr O} \right)
\,\,\, ,\label{3.2}\end{equation}
where
$O$ $(I$) refer to vectors of
outgoing (incoming) plane-wave amplitudes on the left
and $O^{\prime}$, ($I^{\prime}$) to corresponding amplitudes on the right,
 each plane-wave being divided by the
square root of its longitudinal group velocity to ensure unitarity of $s$.
Whereas $T$ connects plane wave amplitudes in the left lead to
amplitudes in the
right lead, the $s$-matrix connects incoming amplitudes to outgoing
amplitudes and satisfies
\begin{equation}\left(\matrix{O \cr O^\prime}\right) = s\left(\matrix{I \cr I^\prime}\right)
\label{3.1}\end{equation}
\noindent
Once $T$ is known, the s-matrix
can be constructed. Indeed if $s$ is written as

\begin{equation}s =
\left(\matrix{r & t^\prime \cr
t & r^\prime }\right) \,\,\, ,\label{3.3}\end{equation}
then $T$ has the form

\begin{equation}
T  =
\left(\matrix{T_{11} & T_{12} \cr
T_{21} & T_{22} }\right)
 =
\left(\matrix{ (t^\dagger)^{-1} & r^\prime(t^\prime)^{-1} \cr
-(t^\prime)^{-1}r & (t^\prime)^{-1} }\right)
\,\,\, ,\label{3.4}\end{equation}
from which the following inverse relation is obtained,

\begin{equation}s =
\left(\matrix{-T_{22}^{-1}T_{21} & T^{-1}_{22} \cr
(T^\dagger_{11})^{-1} & T_{12}T^{-1}_{22} }\right) \,\,\, .\label{3.5}\end{equation}

An alternative method of evaluating coefficients in the above current-voltage
relations is provided by
the recursive Green's function method, which uses Gaussian
elimination to compute the Green's function
on sites located at the surface of the interface between
external normal leads and the scatterer.
Given the surface Green's function, scattering coefficients can be
obtained from generalised Fisher-Lee relations [Fisher and Lee 1982],
derived by [Takane and Ebisawa 1992(a)] and later
rederived in [Lambert 1993, Lambert,
Hui and Robinson 1993].
This recursive technique is identical
to the \lq\lq decimation" method employed by [Lambert and Hui 1990]
and is essentially an efficient implementation of Gaussian elimination.

A third method of evaluating the coefficient $R_a$ was derived
by [Beenakker 1992] for the case where there is perfect Andreev reflection
at the boundary of a clean superconductor. The resulting formula
expresses $R_a$ in terms
of scattering properties of the normal state, as discussed in section V
below and facilitates the application of random matrix theory to
N-S structures [Beenakker 1997].

% multiple scattering methods and numerical methods

\section{Zaitsev's boundary conditions and multiple scattering techniques.}

In the previous section,  the multiple scattering approach to dc transport
was reviewed and in the following section the method of quasi-classical
Green's function will be discussed.
Whereas the multiple scattering method is a stand-alone approach,
the quasi-classical technique must be supplemented by boundary conditions
describing quasi-particle scattering at the interface
between different metals. These boundary conditions are provided
by scattering theory and therefore provide a bridge between the two approaches.
In the literature, quasi-classical boundary conditions have
 not been discussed using
 the  language of modern scattering theory and as a consequence
are restricted to planar surfaces with translational invariance
in the transverse direction. In this section we offer a
rederivation of
Zaitsev's boundary conditions [Zaitsev 1984],
 which not only fills this gap,
but also yields a more general condition applicable
to non-planar interfaces.

As noted in section 2, the starting point for a multiple scattering description
is the scattering matrix $s$, with matrix elements $s_{nn'}$
connecting incoming to outgoing scattering
channels, and satisfying equation (\ref{3.1}).

Clearly $s(E, H)$ is a functional of all physical potentials
entering the Hamiltonian $H$,
as well as a function of E. Since $H$ is Hermitian, quasi-particle probability
is conserved, which yields
\begin{equation}s^{-1}(E, H) = s^{\dagger}(E, H)\label{3.6}\end{equation}
and
\begin{equation}\left(\matrix{1&0\cr0&-1}\right)=
T(E,H)^\dagger\left(\matrix{1&0\cr0&-1}\right)T(E,H)
\label{3.7}.\end{equation}
Furthermore time reversal symmetry yields
\begin{equation}s^*(E, H^*)=s^{-1}(E,H)\label{3.8}\end{equation}
and
\begin{equation}T(E,H)=\left(\matrix{0&1\cr1&0}\right)T^*(E,H^*)\left(\matrix{0&1\cr1&0}\right).
\label{3.9}\end{equation}
Hence
\begin{equation}s(E, H^*)=s^t(E, H)\label{3.10}\end{equation}
and
\begin{equation}
{T^{-1}}^t(E,H^*)=
\left(\matrix{0&1\cr-1&0}\right)T(E,H)\left(\matrix{0&-1\cr1&0}\right)
\label{3.11}.\end{equation}

It should be noted that
the the unitary matrix $s$ in equation (\ref{3.1})
and the matrix $T$ in equation (\ref{3.2})
connects open channels to open channels only.
If evanescent states which decay at large distances from the scatterer
are included in the outgoing states on the left-hand-side of (\ref{3.1})
and states which grow at large distances are included on the right,
then the corresponding scattering matrix $\hat s$ and
 transfer matrix $\hat T$ will not satisfy equations (\ref{3.6}) and
(\ref{3.7}). The matrices $s$ and $T$ should then be constructed by
eliminating closed channels from $\hat s$ and $\hat T$.

Using the above definitions, the boundary conditions derived by Zaitsev
are readily extended to the case of multiple scattering channels.
The problem to be solved is that of connecting the Green's function
$G_j$ in a region $j$ to the Green's function(s) in all other
regions connected to the scatterer.
Although the following analysis is easily generalized to many
leads attached to a scattering region,
for simplicity of notation we restrict the discussion
to only two leads. 

For the purpose of this section, it is convenient to write the channel
index $n$ in the form $n=j,\sigma,p$, where $j=1,2$ labels a lead,
$p$ labels all tranverse quantum numbers, $\sigma =+1$
for a right-going (or right-decaying) wave and $\sigma=-1$ for a
left-going (or left-decaying wave).
If $\underline r_j$ denotes a position in
a crystalline lead $j$ of
constant cross-section, then a plane wave of unit flux
incident along channel $n$
can be written
\begin{equation}\phi_{n}(\underline r_j)=
\chi_{n}(\underline{\rho}_j)\,(v_{n}(E))^{-1/2}
{\rm exp}(\imath k^p_{j\sigma}(E)
z_j)\label{3.12},\end{equation}
 where $\chi_n(\underline{\rho}_j)$ is the transverse mode
 associated with channel $n$
 and
$(v_{n}(E))^{-1/2}{\rm exp}(ik^p_{j\sigma} (E)
z_j)$ is a plane wave of unit flux, with longitudinal wavevector $k^p_{j\sigma}
(E)$.

Similarly for $\underline r_i \ne \underline r_j$,
and $\underline r_i$, $\underline r_j$ located within the leads,
the Green's function
has the form
\begin{equation}
G(\underline {r_i}, \underline r_j)=
\sum_{p,p',\sigma,\sigma'} \chi_{i,\sigma,p}(\underline{\rho}_i)
\chi_{j,\sigma',p'}(\underline{\rho}_j) A^{ij}_{\sigma,\sigma'}(p,p',\Sigma)
{{{\rm exp}\imath[ k^p_{i,\sigma}(E) z_i+k^{p'}_{j,\sigma'}(E)z_j]
}\over{\sqrt{v_{(i,\sigma,p)}(E)v_{(j,\sigma',p')}(E)}}},
\label{3.13}\end{equation}
where $\Sigma$ is equal to the sign of $(z_i-z_j)$.

For $\underline r_i \ne \underline r_j$ and a fixed value of $\Sigma$,
the Green's function is simply
a wavefunction and therefore
the matrix $A^{ij}_{\sigma,\sigma'}$ with matrix elements
$A^{ij}_{\sigma,\sigma'}(p,p',\Sigma)$ satisfy relations of the form
of equation (\ref{3.2}).
First consider the form of the Green's
function when $z_j< z_i$ and $j=1$.
When viewed as a function of $z_i$, for $z_j< 0$,
one has

\begin{equation}
{\hat T}(E,H)\left(\matrix{A^{11}_{++}&A^{11}_{+-}\cr
A^{11}_{-+}&A^{11}_{--}}\right)=
\left(\matrix{A^{21}_{++}&A^{21}_{+-}\cr
A^{21}_{-+}&A^{21}_{--}}\right)
\label{3.14}\end{equation}

On the other hand, when viewed as a function of $z_j$, using the conjugate
equation for the Green's function (which involves the time reversed Hamiltonian,
one obtains for $i=2$ and $z_i>0$,
\begin{equation}
\left(\matrix{A^{22}_{++}&A^{22}_{-+}\cr
A^{22}_{+-}&A^{22}_{--}}\right)=
{\hat T}(E,H^*)\left(\matrix{A^{21}_{++}&A^{21}_{-+}\cr
A^{21}_{+-}&A^{21}_{--}}\right)
\label{3.15}\end{equation}

Hence after eliminating the off-diagonal terms,
$A_{++}^{21}$, etc., we obtain the general
boundary condition relating the  Green's functions
on the left of the scatterer,
to the Green's functions
on the right:
\begin{equation}
{\hat T}(E,H)\left(\matrix{A^{11}_{++}&A^{11}_{+-}\cr
A^{11}_{-+}&A^{11}_{--}}\right)=
\left(\matrix{A^{22}_{++}&A^{22}_{+-}\cr
A^{22}_{-+}&A^{22}_{--}}\right){\hat T}^t(E,H^*)^{-1}
\label{3.16}.\end{equation}

Equation (\ref{3.16}) is a generalisation of Zaitsev's boundary condition to
the case of a  non-planar barrier, which in general may contain impurities
and break time-reversal symmetry.
\footnote{A similar argument with $z_j > z_i$ yields an identical result
and therefore the boundary condition is independent of the choice of
$\Sigma$.}

Since the sum on the right-hand-side of equation (\ref{3.13}) includes evanescent
channels, the general boundary condition involves the matrix $\hat T$.
However at large distances from the scatterer, evanescent channels
can be ignored and therefore $\hat T$ can be replaced by $T$.
In view of equation (\ref{3.11}),
if time-reversal symmetry is present,
 the boundary condition simplifies to
\begin{equation}
T(E,H)\left(\matrix{-A^{11}_{+-}&A^{11}_{++}\cr
-A^{11}_{--}&A^{11}_{-+}}\right)=
\left(\matrix{-A^{22}_{+-}&A^{22}_{++}\cr
-A^{22}_{--}&A^{22}_{-+}}\right)T(E,H)
\label{3.17}\end{equation}

In the work of Zaitsev, where $p$ represents the transverse momentum
and all matrices are diagonal, equation (\ref{3.17})
is satisfied for each separate $p$ and in the notation of Zaitsev
has the form
\begin{equation}
T(E,H)\left(\matrix{-\check g^1_+&\check{\cal G}^1_+\cr
-\check{\cal G}^1_-&\check g^1_-}\right)=
\left(\matrix{-\check g^2_+&\check{\cal G}^2_+\cr
-\check{\cal G}^2_-&\check g^2_-}\right)T(E,H)
\label{3.18}\end{equation}

As an example, taking the trace of equation (\ref{3.18}) yields
\footnote{More precisely, one has to multiply eq. (\ref{3.18}) by
$T^{-1}$ and exploit the ciclyc property of the trace.}
\begin{equation}\check g_+^1-\check g_-^1=\check g_+^2-\check g_-^2\label{3.19},\end{equation}
which demonstrates that the anti-symmetric part of the quasi-classical Green's
function is continuous across a scattering region
\footnote{As it will become clear in the next section, this
corresponds to the conservation of the current}.
More generally,
equation (\ref{3.17}) yields the corresponding relation
\begin{equation}{\rm Tr}(A^{11}_{-+}-A^{11}_{+-})={\rm Tr}(A^{22}_{-+}-A^{22}_{+-})
\label{3.20}.\end{equation}

Similarly generalisations of all other boundary conditions can be expressed
as traces over the matrices $A^{ij}_{\sigma\sigma'}$.
Indeed multiplying both sides of (\ref{3.17}) by each of the matrices

\begin{equation}
\left(\matrix{0 & 0 \cr 0 & T^{-1}_{22} }\right),\,\,\,
\left(\matrix{T^{-1}_{11} & 0 \cr 0 & 0 }\right), \,\,\,
\left(\matrix{0 & 0 \cr T^{-1}_{12} & 0 }\right), \,\,\,
\left(\matrix{0 & T^{-1}_{21} \cr 0 & 0 }\right)
\end{equation}

and taking the trace of the resulting four equations yields
\begin{equation}
{\rm Tr}[A^{11}_{-+}-A^{22}_{-+}]=
-{\rm Tr}[T^{-1}_{22}T_{21}A^{11}_{++}+T_{12}T^{-1}_{22}A^{22}_{--}]
={\rm Tr}[rA^{11}_{++}-r'A^{22}_{--}]
\label{3.21}\end{equation}
\begin{equation}
{\rm Tr}[A^{11}_{+-}-A^{22}_{+-}]=
-{\rm Tr}[T^{-1}_{11}T_{12}A^{11}_{--}+T_{21}T^{-1}_{11}A^{22}_{++}]
={\rm Tr}[r^\dagger A^{11}_{--}-{r'}^\dagger A^{22}_{++}]
\label{3.22}\end{equation}
\begin{equation}
{\rm Tr}[A^{11}_{-+}+A^{22}_{+-}]=
{\rm Tr}[-T^{-1}_{12}T_{11}A^{11}_{++}+T_{22}T^{-1}_{12}A^{22}_{++}]
={\rm Tr}[{r^\dagger}^{-1}A^{11}_{++}+{r'}^{-1}A^{22}_{++}]
\label{3.23}\end{equation}
\begin{equation}
{\rm Tr}[A^{11}_{+-}+A^{22}_{-+}]=
{\rm Tr}[-T^{-1}_{21}T_{22}A^{11}_{--}+T_{11}T^{-1}_{21}A^{22}_{--}]
={\rm Tr}[r^{-1}A^{11}_{--}+{r^{\prime\dagger}}^{-1}A^{22}_{--}]
\label{3.24}\end{equation}

Subtracting (\ref{3.21}) from (\ref{3.22}) yields, in view of (\ref{3.20}),
\begin{equation}
 {\rm Tr}[rA^{11}_{++}-r^\dagger A^{11}_{--}]=
 {\rm Tr}[r'A^{22}_{--}
-{r'}^\dagger A^{22}_{++}]
\label{3.25}\end{equation}
and subtracting (\ref{3.23}) from (\ref{3.24}) yields
\begin{equation}
{\rm Tr}[{r^\dagger}^{-1}A^{11}_{++}-r^{-1}A^{11}_{--}]
={\rm Tr}[{r^{\prime\dagger}}^{-1}A^{22}_{--}-{r'}^{-1}A^{22}_{++}]
\label{3.26}\end{equation}

Adding (\ref{3.21}) to (\ref{3.22}) and (\ref{3.23}) to (\ref{3.24}) yields
\begin{equation}
{\rm Tr}[A^{11}_{+-}+A^{11}_{-+}]-
{\rm Tr}[A^{22}_{+-}+A^{22}_{-+}]=
{\rm Tr}[rA^{11}_{++} +r^\dagger A^{11}_{--}]
-{\rm Tr}[r'A^{22}_{--}+
{r'}^\dagger A^{22}_{++}]
\label{3.27}\end{equation}
and
\begin{equation}
{\rm Tr}[A^{11}_{+-}+A^{11}_{-+}]+
{\rm Tr}[A^{22}_{+-}+A^{22}_{-+}]
={\rm Tr}[{r^\dagger}^{-1}A^{11}_{++}+r^{-1}A^{11}_{--}]
+{\rm Tr}[{r'}^{-1}A^{22}_{++}
+{r^{\prime\dagger}}^{-1}A^{22}_{--}]
\label{3.28}.\end{equation}

For the translationally invariant case considered by Zaitsev,
writing
$t=t'=\vert t\vert\exp\imath\theta$ $r=\vert r\vert\exp\imath\phi$,
$r'=-\vert r\vert\exp\imath(2\theta-\phi)$,
$A^{ii}_{\sigma\sigma}=\check{\cal G}^j_\sigma$ and
$A^{ii}_{\sigma -\sigma}=\check g^j_\sigma$, one obtains
from either of equations (\ref{3.25}) and (\ref{3.26})
\begin{equation}\check{\cal G}_a^1=\check{\cal G}_a^2\label{3.29},\end{equation}
from (\ref{3.27}) and (\ref{3.28}),
\begin{equation}\check g^1_s-\check g^2_s=\vert r\vert(\check{\cal G}^1_s+\check{\cal G}^2_s)
\label{3.30},\end{equation}
\begin{equation}\check g^1_s+\check g^2_s={{1}\over{\vert r\vert}}(\check{\cal
 G}^1_s-\check{\cal G}^2_s)\label{3.31},\end{equation}
and from
(\ref{3.19})
\begin{equation}\check g_a^1=\check g_a^2\label{3.32},\end{equation}
where
\begin{equation}\check{\cal G}_{s,a}^1=[\check{\cal G}^1_+\exp\imath\phi\pm \check{\cal
G}^1_-\exp-\imath\phi]/2,
\end{equation}
\begin{equation}\check{\cal G}_{s,a}^2=[\check{\cal G}^2_+\exp\imath(\phi-2\theta)
\pm \check{\cal G}^2_-\exp\imath(2\theta-\phi)]/2
\end{equation}
and
\begin{equation}\check g^i_{s,a}=[\check g^i_+\pm \check g^i_-]/2.\end{equation}

Equations (\ref{3.29}) and (\ref{3.32}), which are a limiting case of the more general
relation (\ref{3.17}) are identical to the boundary conditions of
Zaitsev and will be discussed in the next section.

% Quasi-classical boundary conditions
%\documentstyle{article}
%\begin{document}
\section{Quasi-classical Green's function approach}
\subsection{Quasi-classical equations}

The quasi-classical approach to the theory of superconductivity, 
initiated by Eilenberger [1968] and Larkin and Ovchinnikov [1968] and
developed by several authors [Usadel 1970, Eliashberg 1971, Larkin
and Ovchinnikov 1973,1975] has been largely used to
analyse  transport phenomena in dirty hybrid systems.
This approach can be used to study thermodynamic and kinetic properties
of superconductors, whose dimensions significantly exceed the 
Fermi wavelength $\lambda_F =2\pi /k_F$.
For the purpose of reviewing recent theoretical work,
 we briefly summarize the
quasi-classical Green's function approach,
 following mainly
[Larkin et Ovchinnikov 1986].
There exist in the literature several excellent reviews on
the subject including [Rammer and
Smith 1986] and we refer the reader to these for a more detailed
exposition. The derivation of the equation for quasi-classical
 Green's function starts from the
Dyson equation for the matrix Green's function
 ${\check G}$ in the non equilibrium Keldysh formalism
[Keldysh 1964)].

\begin{equation}
({\check G}^{-1}_0-{\check {\Sigma}}){\check G}={\check 1}
\label{4.1}
\end{equation}

where

\begin{equation}
{\check G}=\left(\begin{array}{c c}
{\hat G}^R & {\hat G} \\
0 & {\hat G}^A\\
\end{array}
\right).
\label{4.2}
\end{equation}

Following standard notation, we indicate with a
 "hat" matrices  in Nambu space and define
 retarded, advanced and Keldysh Green's functions

\begin{equation}
{\hat G}^R (1,2) =\theta (t_1-t_2) ({\hat  G}^{-+} (1,2)-
{\hat G}^{+-} (1,2))
\label{4.5}
\end{equation}

\begin{equation}
{\hat G}^A (1,2) =-\theta (t_2-t_1) ({\hat  G}^{-+} (1,2)-
{\hat G}^{+-} (1,2))
\label{4.6}
\end{equation}

\begin{equation}
{\hat G} (1,2) = {\hat  G}^{-+} (1,2)+
{\hat G}^{+-} (1,2).
\label{4.7}
\end{equation}

where 
\begin{equation}
i{\hat G}^{-+}=\left(\begin{array}{c c}
<\psi_{\uparrow}(1)\psi^{\dagger}_{\uparrow}(2)> &
<\psi_{\uparrow}(1)\psi_{\downarrow}(2)> \\
-<\psi_{\downarrow}^{\dagger}(1)\psi^{\dagger}_{\uparrow}(2)> &
-<\psi_{\downarrow}^{\dagger}(1)\psi_{\downarrow}(2)>\\
\end{array}
\right)
\label{4.3}
\end{equation}

\begin{equation}
i{\hat G}^{+-}=-\left(\begin{array}{c c}
<\psi_{\uparrow}^{\dagger}(2)\psi_{\uparrow}(1)> &
<\psi_{\downarrow}(2)\psi_{\uparrow}(1)> \\
-<\psi_{\uparrow}^{\dagger}(2)\psi^{\dagger}_{\downarrow}(1)> &
-<\psi_{\downarrow}(2)\psi^{\dagger}_{\downarrow}(1)>\\
\end{array}
\right).
\label{4.4}
\end{equation}

In these expressions,
 $1\equiv ({\bf r}_1 , t_1 )$ and $2\equiv ({\bf r}_2 , t_2 )$,
which allow equation (\ref{4.1}) to be written in
the less compact form

$$
\int d2 ({\check G}_0^{-1} (1,2)-{\check {\Sigma}}(1,2)){\check G}(2,3)=
{\check {\delta}}(1-3).
$$

The equation  conjugate to equation (\ref{4.1}) is

\begin{equation}
{\check G}({\check G}^{-1}_0-{\check {\Sigma}})={\check 1}
\label{4.8}
\end{equation}

and  taking the difference between equations(\ref{4.1}) and (\ref{4.8})
yields

\begin{equation}
\left[ {\check G}_0^{-1}-{\check {\Sigma}}, {\check G}\right]=0.
\label{4.9}
\end{equation}
This equation is simplified
by going to the  center-of-mass and relative coordinates 
(${\bf R},T$) and (${\bf r},t$) defined as 

$$
{\bf r}_{1,2}={\bf R}\pm {\bf r}/2,~~~t_{1,2}=T\pm t /2,
$$

 Fourier transforming with respect to ${\bf r}$ and $t$, and
  introducing the quasi-classical Green's function defined by

\begin{equation}
{\check g}({\bf R}, T; {\hat p},\epsilon )= {i\over \pi}
\int_{-\infty}^{\infty} d\xi {\check G} ({\bf R}, T; {\bf p},\epsilon )
\label{4.10}
\end{equation}
where $\xi =p^2/2m -\mu$ is the energy measured with respect to the
Fermi level
and ${\hat p}$ is the unit vector in the direction of the momentum ${\bf p}$. 
 On the assumption that
  the self-energy depends weakly on the energy  $\xi$, 
one sets $\xi =0$ in ${\check {\Sigma}}$, which yields

\begin{equation}
\partial_T \lbrace {\check {\tau}}_z , {\check g}\rbrace +v_F
{\hat p} \cdot \partial_{\bf R} {\check g} -i\epsilon
\left[ {\check {\tau}}_z , {\check g}\right] +
i\left[ {\check {\Sigma}}, {\check g}\right]=0.
\label{4.11}
\end{equation}

where the curly brackets denote the anticommutator.
 This is the equation for the quasiclassical Green's function first
derived by Eilenberger [1968]. 
${\check {\tau}}_z$ is a block-diagonal matrix with diagonal
block entries ${\hat {\sigma}}_z$.
In deriving
equation (\ref{4.8}),  the non-homogenous term present in
 the Dyson equation (\ref{4.1}) was eliminated and
as a consequence, equation (\ref{4.11})
determines ${\check g}$ only up to a constant.
 As discussed
 in  [Shelankov 1985, Zaitsev 1984],
a useful normalization
 condition for ${\check g}$ is

\begin{equation}
{\check g}{\check g}=1.
\label{4.12}
\end{equation}

 In terms of the quasiclassical
Green's function, the physical current is given by 

\begin{equation}
\label{4.13}
{\bf j}({\bf R}, T)=-{1\over 4}N_0v_F\int^{\infty}_{\infty}
 d\epsilon < \mu
Tr(\sigma_z {\hat g} (\epsilon, {\hat p}, {\bf R}, T))>,
\end{equation}

where $N_0=m^2v_F/2\pi^2$ is the free single-particle density of states
per spin, $\sigma_z$ the
usual Pauli matrix and $v_F$ the Fermi velocity.
 Here $<...>$ indicates the  average over the angle $\theta$
 formed by ${\hat p}$
and the direction of ${\bf R}$, i.e.,
$(1/2)\int^1_{-1}d\cos (\theta ) (...)$. 
It should be noted that, because of the angular average,
 only the antisymmetric
part of ${\hat g}$  enters the expression for the current.

In the dirty limit, in the case of an isotropic scattering
 potential with elastic scattering time $\tau$,
the effect of   non-magnetic impurities
can be described by the self-energy

\begin{equation}
{\check {\Sigma}} =-{{i}\over {2\tau}} < {\check g}>\,\, ,
\label{4.14}
\end{equation}
 and
equation (\ref{4.11}) can be considerably simplified
[Usadel 1970].
In this case one expands ${\check g}$ in spherical harmonics keeping
only the s- and p-wave terms

\begin{equation}
{\check g} (\cos (\theta ) )={\check g}_0 + \cos (\theta ) {\check g}_1
\label{4.15}
\end{equation}
with ${\check g}_0$ and ${\check g}_1$ not depending on
$cos (\theta ) $ and
$cos (\theta ) {\check g}_1 \ll {\check g}_0$.
By inserting equation (\ref{4.15}) in equation (\ref{4.11}),
${\check g}_1$ is expressed in terms of ${\check g}_0$

\begin{equation}
{\check g}_1 =-l {\check g}_0 \partial_{\bf R} {\check g}_0
\label{4.16}
\end{equation}
and for ${\check g}_0$ one obtains a diffusion-like equation

\begin{equation}
D\partial_{\bf R} {\check g}_0\partial_{\bf R} {\check g}_0 +i\epsilon
\left[ {\check {\tau}}_z , {\check g}_0 \right]
-\partial_T \lbrace {\check {\tau}}_z , {\check g}_0 \rbrace=0
\label{4.17}
\end{equation}
with $D=v_F l\tau/3$  the diffusion coefficient and
$l=v_F\tau$ the mean free path. 

Eqs.(\ref{4.11}) and (\ref{4.17})
are valid for  bulk systems and in the
 presence of boundaries, must be supplemented by
boundary conditions which connect the Green functions evaluated in
different regions. These boundary conditions have
turned out to be crucial to the recent development  of transport
theory in hybrid structures. For this reason, we
present a brief discussion of these  in the following subsection.

\subsection{Boundary conditions for quasiclassical Green functions}

In this subsection,  we  discuss the boundary conditions for the
quasiclassical Green's function [Zaitsev 1984], given by equations
(\ref{3.29}) to (\ref{3.32}) of section III.
The antisymmetric functions ${\check g}_a=
{\check g}_a^1={\check g}_a^2$ and
${\check {\cal G}}_a={\check {\cal G}}_a^1={\check {\cal G}}_a^2$
are continuous across the boundary, while
the symmetric ones ${\check g}_s^i$ and ${\check {\cal G}}_s^i$ 
 experience a jump determined by the transparency of the barrier. 
The size of the jump vanishes for perfectly transmitting interfaces.

The function ${\check g}$ is the quasi-classical
Green's function satisfying equation (\ref{4.11}), whereas
the Green's function ${\check {\cal G}}$ arises
from reflected waves at the boundary.
It is therefore necessary to derive an equation of
motion for ${\check {\cal G}}$ to be used together
 with equation (\ref{4.11}).  We will
not show this derivation here, because in actual calculations,
one only uses the Green's function ${\check g}$. Zaitsev has shown,
that the final result is of the form
\begin{equation}
{\check g}^i (k_i ){\check {\cal G}}^i(k_i )=(-1)^i sgn(k_i)
 {\check {\cal G}}^i (k_i ),
\label{4.46}
\end{equation}
which must be used in conjunction with the normalization condition

\begin{equation}
{\check g}^i {\check g}^i = {\check 1}.
\label{4.47}
\end{equation}

The  boundary conditions for the antisymmetric components
are already  decoupled with respect to the
${\check g}_a$ and ${\check {\cal G}}_a$. To decouple the
symmetric components, we express
 eqs.(\ref{4.46}-\ref{4.47}) 
in terms of the symmetric and antisymmetric parts to yield

\begin{equation}
\begin{array}{c}
{\check g}^i_s{\check {\cal G}}^i_s +{\check g}_a{\check {\cal G}}_a
=(-1)^i{\check {\cal G}}_a\\
{\check g}^i_s{\check {\cal G}}_a +{\check g}_a{\check {\cal G}}^i_s
=(-1)^i{\check {\cal G}}^i_s\\
\end{array}
\label{4.48}
\end{equation}

and

\begin{equation}
{\check g}^i_s{\check g}^i_s+{\check g}_a{\check g}_a={\check 1},~~
{\check g}^i_s{\check g}_a+{\check g}_a{\check g}^i_s=0.
\label{4.49}
\end{equation}

By manipulating equation (\ref{4.48}) one  obtains

\begin{equation}
{\check g}^1_s {\check {\cal G}}^1_s +{\check g}^2_s {\check {\cal G}}^2_s=
{\check g}_a 
({\check g}^1_s {\check {\cal G}}^1_s-{\check g}^2_s {\check {\cal G}}^2_s)
\label{4.50}
\end{equation}

which is  the extra condition to be used together with
equations (\ref{3.30}-\ref{3.31}).
By means of equations (\ref{3.30}-\ref{3.31}) one expresses ${\check {\cal G}}^2_s$
and ${\check {\cal G}}^1_s$ in terms of ${\check g}_s^1$ and 
${\check g}_s^2$ and substitute in equation (\ref{4.50}) so that
 the final boundary condition
reads

\begin{equation}
{\check g}_a \left[ R(1-{\check g}_a{\check g}_a)+(T/4)({\check g}_s^1-
{\check g}^2_s)^2\right]=(T/4)({\check g}_s^2{\check g}_s^1-{\check g}_s^1
{\check g}_s^2)
\label{4.51}
\end{equation}

where $R=|r|^2$, $T=1-R$ are the reflection and transmission coefficient
of the barrier.

As a simple example, we consider   the
normal case, when the Green's function is  a two-by-two matrix

\begin{equation}
{\check g}=\left(
\begin{array}{c c}
1 & g \\
0 &-1 \\
\end{array}
\right)
\label{4.52}
\end{equation}

with $g=2f$, $f$ being the usual distribution function
entering the Boltzman kinetic equation.  By observing that

\begin{equation}
[{\check g}_{1s},{\check g}_{2s}]=4\left(
\begin{array}{c c}
0 & f_{1s} -f_{2s} \\
0 &0 \\
\end{array}
\right);~~
({\check g}_1-{\check g}_2)^2={\check 0};~~
{\check g}_{a}^2={\check 0}
\label{4.53}
\end{equation}

 the boundary condition (\ref{4.51}) assumes the form

\begin{equation}
f_a=(T/2R)(f_{1s}-f_{2s}).
\label{4.54}
\end{equation}

It is worth noticing that this result could have been obtained
directly from simple
 counting arguments of the form

\begin{equation}
f_1(k)=Tf_2(k)+Rf_1(-k),~~~
f_2(-k)=Tf_1(-k)+Rf_2(k)
\label{4.55}.
\end{equation}

In view of equation (\ref{4.13}) for the current, equation (\ref{4.54})
yields for the conductance of a tunnel junction in the normal case,

\begin{equation}
\label{4.56}
G_T=e^2N_0v_F <\mu T/R>
\end{equation}

where $\mu =\theta$.
In the general case, equation (\ref{4.51}) can be simplified in the low transparency
limit.
If $T\ll 1$, then ${\check g}_a \approx T$ and the boundary
condition equation (\ref{4.51}) reduces
to

\begin{equation}
{\check g}_a =(T/4R) [{\check g}_{1s}, {\check g}_{2s} ].
\label{4.57}
\end{equation}

In the dirty limit, ${\check g}_a =-\mu l{\check g}_s\partial_R
{\check g}_s$ and equations (\ref{4.16}-\ref{4.17}) yield the
boundary condition [Kuprianov and Lukichev 1988]

\begin{equation}
l{\check g}_{1s}\partial_R {\check g}_{1s}=
l{\check g}_{2s}\partial_R {\check g}_{2s},
\label{4.58}
\end{equation}

which after multiplying equation (\ref{4.57}) by $\mu$ and taking the angular
average yields

\begin{equation}
l{\check g}_{1s}\partial_R {\check g}_{1s}={3\over 4}
<\mu T/R> [{\check g}_{2s}, {\check g}_{1s} ].
\label{4.59}
\end{equation}

By defining a  conserved "super" current $\check {I}$ (cf. equation (\ref{4.13}))
one finally arrives at the following boundary condition

\begin{equation}
{\check I} = {{\sigma}\over e}
 {\check g}_{2s} \partial_{\bf R} {\check g}_{2s} 
= {{G_T}\over {2e}} [ {\check g}_{2s} , {\check g}_{1s} ],
\label{4.61}
\end{equation}

where $\sigma = 2e^2N_0 v_F l/3$ is the Drude electrical conductivity.
Equation (\ref{4.61}) is the desired boundary condition to be used together
with the diffusion equation
(\ref{4.17}) in the presence of boundaries.
Eq.(\ref{4.61}) is strictly valid in the case of small barrier transparency.
Such restriction has been recently somehow relaxed by Lambert, Raimondi,
Sweeney and Volkov [1997].
In the next subsection we show how to
compute the current-voltage relation of the two fundamental elements
 of an hybrid system, namely a tunnel junction and a diffusive region.

\subsection{Quasi-classical theory at work.}
\noindent
{\bf (i) A Tunnel junction.}

Consider  the current through a tunnel junction as
given by the commutator on the right-hand side of equation (\ref{4.61}).
The physical current is obtained from the Keldysh component of 
equation (\ref{4.61})

$$
[ {\check g}_2 , {\check g}_1 ]_k = 
{\hat g}_2^R {\hat g}_1+{\hat g}_2{\hat g}_1^A- 
{\hat g}_1^R{\hat g}_2-{\hat g}_1{\hat g}_2^A
$$

where we have dropped   the subscript "s".
The normalization condition, ${\check g} {\check g} = {\check 1}$, 
allows us  to choose

$$
{\hat g}_{1,2} =
 {\hat g}_{1,2}^R {\hat f}_{1,2}-{\hat f}_{1,2}{\hat g}_{1,2}^A
$$

where the matrix ${\hat f}$ can be taken to be diagonal
[Larkin and  Ovchinnikov 1975]

$$
{\hat f}_{1,2} =f^0_{1,2} {\hat \sigma}_0 +f^z_{1,2} {\hat \sigma}_z.
$$

As a result,  multiplying by ${\hat\sigma}_z$ and taking the trace,
yields (cf. equation (\ref{4.13}))

\begin{equation}
j={{G_T}\over {16e}}\int^{\infty}_{-\infty}d\epsilon
Tr ({\hat\sigma}_z ( f^0_1 {\hat I}_a + f^0_2 {\hat I}_b +
f^z_1 {\hat I}_c + f^z_2 {\hat I}_d))
\label{4.62}
\end{equation}

where
$$
{\hat I}_a =
\left[ {\hat g}_2^R ({\hat g}_1^R-{\hat g}_1^A )-
({\hat g}_1^R-{\hat g}_1^A ){\hat g}_2^A\right]
$$

$$
{\hat I}_b =-
\left[ {\hat g}_1^R ({\hat g}_2^R-{\hat g}_2^A )-
({\hat g}_2^R-{\hat g}_2^A ){\hat g}_1^A\right]
$$

$$
{\hat I}_c =
\left[ {\hat g}_2^R ({\hat g}_1^R{\hat \sigma}_z -{\hat \sigma}_z{\hat g}_1^A )-
({\hat g}_1^R{\hat \sigma}_z-{\hat \sigma}_z{\hat g}_1^A ){\hat g}_2^A\right]
$$

$$
{\hat I}_d =-
\left[ {\hat g}_1^R ({\hat g}_2^R{\hat \sigma}_z-{\hat \sigma}_z{\hat g}_2^A )-
({\hat g}_2^R{\hat \sigma}_z-{\hat \sigma}_z{\hat g}_2^A ){\hat g}_1^A\right].
$$

Due to the normalization condition
 ${\hat g}^{R(A)}{\hat g}^{R(A)}={\hat 1}$,
we have
$$
{\hat g}^{R(A)} = {\bf g}^{R(A)} \cdot {\bf \sigma}\equiv
\sum_{i=1}^3 g^{R(A)}_i \sigma_i
$$

where ${\bf g}^{R(A)}= (iF^{R(A)}sin(\phi )$,
$iF^{R(A)}cos(\phi ), G^{R(A)})$ and
$\phi$ is the phase of the superconducting order parameter.
Hence

\begin{equation}
j={{G_T}\over {8e}}\int^{\infty}_{-\infty}d\epsilon
 (I_J + I_{PI})
\label{4.63}
\end{equation}

where

$$
I_J = i  sin(\phi_1 -\phi_2 ) \left[ f^0_2 ( F^R_2 -F^A_2)
 ( F^R_1 +F^A_1)+ 
f^0_1 ( F^R_2 +F^A_2) ( F^R_1 -F^A_1) \right]
$$

and

$$
I_{PI} = \left[ (G^R_1-G^A_1)(G^R_2-G^A_2) +
cos(\phi_1 -\phi_2 ) ( F^R_2 +F^A_2) ( F^R_1 +F^A_1)\right]
(f^z_1 -f^z_2).
$$

In equation (\ref{4.63}), $I_J$ is the Josephson current, while
 $I_{PI}$ is  sometimes referred to as quasi-particle and interference
 current.
 To clarify equation (\ref{4.63}), consider the case where the two regions
 on the left
and right of the barrier are at equilibrium and $\phi_1 =\phi_2$, so that
the  Josephson current vanishes. If the distribution functions $f^z_{1,2}$
have their equilibrium form,
$$
f^z_{1,2}={1\over 2}( tanh((\epsilon +eV_{1,2})/2T)-
tanh((\epsilon -eV_{1,2})/2T))
$$

then the current through   the junction becomes

\begin{equation}
\label{4.64a}
j={{G_T}\over {8e}}
\int^{\infty}_{-\infty}~d\epsilon (f_1^z -f_2^z ) M_{12}
\end{equation}

where

$$
M_{12}={1\over 4} ((G^R_1-G^A_1)(G^R_2-G^A_2) + 
( F^R_2 +F^A_2) ( F^R_1 +F^A_1)).
$$

At $T=0$, this reduces to

\begin{equation}
I=G_T M_{12}|_{\epsilon =0} (V_1-V_2).
\label{4.64}
\end{equation}

Equation (\ref{4.64}) shows that conductance of the tunnel junction is renormalized
by a term $M_{12}$ which depends on the amount of superconducting pairing on the two sides
of the junction. In the limit of normal systems, $M_{1,2}=1$,
 and one recovers
the conductance of the normal state (cf. equation (\ref{4.56})).

\vskip 0.3cm
\noindent
{\bf (ii) A diffusive region.}
\vskip 0.3cm

We derive now  the equivalent of equation (\ref{4.64a}) for
a diffusive region in the absence of inelastic processes.
In this case the equation for the "hat" components
of the Green's function (\ref{4.17})
 reads

$$
\partial_{\bf R} {\hat g}^{R(A)}\partial_{\bf R} {\hat g}^{R(A)} =
i\epsilon \left[ {\hat\sigma}_z , {\hat g}^{R(A)} \right],~~~
\partial_{\bf R} ( {\hat g}^R\partial_{\bf R} {\hat g}+ 
{\hat g}\partial_{\bf R} {\hat g}^A ) =0.
$$
Using  ${\hat g} = {\hat g}^R {\hat f}-{\hat f}{\hat g}^A$ and the equation for
${\hat g}^{R(A)}$, the equation for the Keldysh component becomes

$$
\partial_{\bf R}   \left[ \partial_{\bf R} {\hat f} -{\hat g}^{R}
(\partial_{\bf R} {\hat f}){\hat g}^{A}\right]-
({\hat g}^R\partial_{\bf R} {\hat g}^R )(\partial_{\bf R} {\hat f})-
(\partial_{\bf R} {\hat f})({\hat g}^A\partial_{\bf R} {\hat g}^A)
=0.
$$

The last equation yields ${\hat f}$, once ${\hat g}^R$ and ${\hat g}^A$ have
been determined. Choosing 
${\hat f}=f^0 {\hat \sigma}_0 +f^z {\hat \sigma}_z$ and
${\hat g}^R= G^R{\hat \sigma}_z +i F^R {\hat \sigma}_y$, 
yields,
after multiplying by ${\hat\sigma}_z$ and taking the trace,

\begin{equation}
\partial_{\bf R} \left[ (1-G^RG^A-F^RF^A)\partial_{\bf R} f^z\right]=0.
\label{4.65}
\end{equation}

If we consider a diffusive region of length $L$, 
the distribution function depends only on the longitudinal
coordinate $x$ and  $f^z (x)$ has the form

\begin{equation}
f^z (x) = m(x) {{f^z(L)-f^z(0)}\over {m(L)}} + f^z (0)
\label{4.66}
\end{equation}

where 
$$
m(x)=\int_0^x~dx' {1\over {1-G^R(x')G^A(x')-F^R(x')F^A(x')}}.
$$

Using the formula for the physical current (\ref{4.13}) one obtains

\begin{equation}
\label{4.66b}
I={{\sigma}\over {8e}} \int_{-\infty}^{\infty} d\epsilon
(f^z(L)-f^z(0)){1\over {m(L)}}.
\end{equation}

To use equation (\ref{4.66b}) one has to solve the equation for ${\hat g}^R$.
Since ${\hat g}^R{\hat g}^R=1$, one can write
$G^R=cosh (u)$ and $F^R=sinh (u)$, which yields

\begin{equation}
D\partial^2_x u+2i\epsilon sinh(u)=0.
\label{4.67}
\end{equation}

Equation (\ref{4.67}) is a non-linear equation, for which
approximate
analytical [Zaitsev 1990, Volkov et al. 1993, Volkov 1994,
Zaitsev 1994] and  numerical  [Zhou et al, 1995, Yip 1995,
Nazarov and Stoof 1996, Golubov et al 1997]
solutions have been obtained.
For the purposes of the present discussion, we note that at zero energy,
the solution may be obtained easily and takes the form

$$
u(x)={{u(L)-u(0)}\over L} x +u(0).
$$

Since

$$
1-G^RG^A-F^RF^A=2cosh^2(Re(u)),
$$

the current (\ref{4.66b}) becomes

\begin{equation}
I={{\sigma }\over L} 
{{tanh(Re(u(L)))-tanh(Re(u(0)))}\over{Re(u(L))-Re(u(0))}}V,
\label{4.68}
\end{equation}

where we have  used the fact that for $T\rightarrow 0$,
${1\over {4e}}\int_{-\infty}^{\infty} d\epsilon(f^z(L)-f^z(0))\rightarrow V$. 
The values $u(L)$ and $u(0)$ are determined by attaching the diffusive region to
electrodes which are assumed to be in equilibrium. For a normal system
$u\equiv 0$, while for a superconducting one $Re(u)=0$ and $Im(u)=-1$. As a result
the resistance of a diffusive region does not depend on the nature of the 
electrodes. 

Nazarov recognized [Nazarov 1994] that the simplicity of the results
of equations (\ref{4.64},\ref{4.68}) can be generalized to an arbitrary
circuit with normal and superconducting electrodes. In fact, the 
evaluation of the conductance can be reduced to the use of a 
set of simple rules, which we now state. A circuit is made of 
dispersive elements (diffusive regions and tunnel junctions)
connected to one another and to the electrodes by nodes.
To each circuit element is associated a "scalar" current, $j$,
which is just the physical  current and a "vector" current
${\bf I}$. To each node is associated a "spectral" unit vector
${\bf s}$. The vector current flowing through a dispersive element
depends on the spectral vectors at the two ends. 
It is useful to imagine the spectral vector as  the spherical
coordinate in the northern hemisphere. The rules are:

I) Conductance in the presence of superconducting electrodes
is the same as in normal circuits, with renormalized tunnel junction
conductances. The renormalization factor is given by the scalar 
product of the spectral vectors at the two sides of the junction.

II) Spectral vector of a normal electrode is at the north
pole ($\hat z$), while that of  a superconducting electrode lies on the
equator, with the longitude given by the phase of the order parameter.

III) Vector current is perpendicular to
both the spectral vectors at the ends of a given element element.
If $\phi$ is the angle between the spectral vectors at the two
ends of a given elements element, the magnitude of the vector current is
$I=G_D \phi$ and $I=G_Tsin(\phi )$ for diffusive and tunnel junction
elements, respectively. 

IV) Vector current is conserved at each node (Generalized
Kirchoff rule).

These rules will be applied to a N-I-S interface
 in the following section.

% solution for an NS interface

%\documentstyle[12pt]{article}
%\begin{document}
\section{The N-S interface and zero-bias anomalies.}
In this section, we show how the multiple-scattering approach
and quasi-classical theory can be used to describe
the simplest example of phase-coherent structure,
namely a N-S interface.

\subsection{ BTK theory and the Andreev approximation}
The simplest problem of a one-dimensional N-I-S
system was examined by
[Blonder, Tinkham, and Klapwijk 1982]. This solution is also
valid in higher dimensions, provided there is translational
invariance in the direction perpendicular to the electronic motion. To see
this, imagine a two-dimensional system with a finite width in y
direction and infinite in the x direction. The motion along y is quantized,
and for fixed $k_y$, the problem becomes one-dimensional and the Bogolubov-de
Gennes equations reduces to

\begin{equation}
\begin{array}{c}
E\psi (x)=(- 
\frac{\hbar ^2}{2m}\partial_x^2-\mu )\psi (x)+\Delta (x)\varphi (x) \\ E\phi
(x)=(\frac{\hbar ^2}{2m}\partial_x^2+\mu )\varphi (x)+\Delta (x)\psi (x)
\end{array}
\label{5.1}
\end{equation}
where $\overline\mu ={\mu }-\hbar^2k_y^2/2m$ is the effective chemical potential in
a given channel. For a pairing potential with a step-like spatial
variation of the form

\begin{equation}
\label{5.2}
\Delta (x)=\Delta _0\Theta (x),
\end{equation}
the Bogolubov-de Gennes equations can be solved by matching
wave functions. For $x<0$  the solution $(\psi ,\phi )$
for an incident electron from the left is

\begin{equation}
\label{5.3}
\begin{array}{c}
\psi ^L(x)=  e^{ikx}+r_0e^{-ikx} \\ 
\phi ^L(x)=  r_ae^{iqx} 
\end{array}
\end{equation}

and for $x>0$
\begin{equation}
\label{5.4}
\begin{array}{cc}
\psi ^R(x)= & t_0ue^{i 
\overline{k}x}+t_a\overline{u}e^{-i\overline{q}x} \\ \phi ^R(x)= & t_0ve^{i 
\overline{k}x}+t_a\overline{v}e^{-i\overline{q}x} .
\end{array}
\end{equation}
The coherence factors $u$ and $v$ correspond to
 a particle-like excitation of
energy $E$ and momentum $\overline{k}$, while $\overline{u}$ and $\overline{v}
$ correspond to a hole-like excitation at the same energy and with momentum
$\overline{q}$, where $E=(\hbar ^2/2m){ k}^2-\overline\mu
=-(\hbar ^2/2m){ q}^2+\overline\mu $
and $E^2=[(\hbar ^2/2m)\overline{k}^2-\overline\mu ]^2+\Delta ^2=[(\hbar ^2/2m)
\overline{q}^2+\overline\mu ]^2+\Delta ^2.$ Also
\begin{equation}
\label{5.5}
\begin{array}{cc}
u^2(v^2)= & \frac 12(1\pm 
\sqrt{E^2-\Delta ^2}/E) \\ \overline{v}^2(\overline{u}^2)= & \frac 12(1\pm 
\sqrt{E^2-\Delta ^2}/E) 
\end{array}
\end{equation}

with $sign(u/v)=sign(\Delta /(E-\epsilon _{\overline{k}}))$, $sign(\overline{
u}/\overline{v})=sign(\Delta /(E-\epsilon _{\overline{q}}))$, where $
\epsilon _{\overline{k}}=(\hbar ^2/2m)\overline{k}^2-\overline\mu $
and $\epsilon _{\overline{q}}=(\hbar ^2/2m)\overline{q}^2+\overline\mu $ for
$E>\Delta $, while for $E<\Delta $ we have 
\begin{equation}
\label{5.6}
\begin{array}{cc}
u/v= & \Delta /(E-i 
\sqrt{\Delta ^2-E^2}) \\ \overline{u}/\overline{v}= & \Delta /(E+i\sqrt{
\Delta ^2-E^2}) .
\end{array} 
\end{equation}

To solve for the scattering coefficients $r_{0\text{, }}r_a$, $t_0$, and $
t_a $ one needs matching conditions at the interface $x=0$. These are the
usual continuity condition for $(\psi ,\phi )$ and their derivatives and
read
\begin{equation}
\label{5.7}
\begin{array}{c}
1+r_0=  t_0u+t_a 
\overline{u} \\ r_a=  t_0v+t_a 
\overline{v} \\ 1-r_0=  \frac{\overline{k}}kt_0u-\frac{\overline{q}}kt_a 
\overline{u}+(1+r_0)U_0 /ik\\ r_a= 
\frac{\overline{k}}kt_0v-\frac{\overline{q}
}kt_a\overline{v}+r_aU_0 /iq
\end{array}
\end{equation}

where the term in $U_0$ in the last two equations of
(\ref{5.7}) allows for a delta-like
potential $U(x)=U_0\delta (x)$ in order to reproduce the
BTK model of an N-I-S interface. The
 system (\ref{5.7}) completely determines the four scattering coefficients. The
detailed solution is of course straightforward, but the resulting
expressions look rather cumbersome, so  it is
convenient to resort to  Andreev's approximation of setting $k=q=\overline{k
}=\overline{q}$, which amounts to neglecting terms of the order $\Delta /E_F.$
Writing $z=2mU_0/2k\hbar ^2$ yields
\begin{equation}
\label{5.8}
\displaystyle
\begin{array}{c}
t_0=\frac{\displaystyle{
\overline{v}(1-iz)}}{\displaystyle{
u\overline{v}(1+z^2)-v\overline{u}z^2}}\\
t_a=\frac{\displaystyle{
ivz}}{\displaystyle{
u\overline{v}(1+z^2)-v\overline{u}z^{21}}}\\
r_a=\frac{\displaystyle{
v\overline{v}}}{\displaystyle{
u\overline{v}(1+z^2)-v\overline{u}z^2}} \\
r_0=\frac{\displaystyle{
(v\overline{u}-u\overline{v})(z^2+iz)}}{\displaystyle{
u\overline{v}(1+z^2)-v 
\overline{u}z^2}} \\
\end{array}
\end{equation}

For $E<\Delta$, where $T_0=T_a=0$, this yields
\begin{equation}
\label{5.9}
R_a=\frac{\Delta ^2}{E^2+(1+2z^2)^2(\Delta ^2-E^2)} 
\end{equation}
and in view of unitarity, $R_0=1-R_a$. For
$E>\Delta$ one
 obtains
\begin{equation}
\label{5.10}
\begin{array}{c}
R_a=\frac{\displaystyle{
\Delta ^2}}{\displaystyle{
(E+(1+2z^2)^2\sqrt{E^2-\Delta ^2}))^2}}\\ 
R_0=\frac{\displaystyle{
4z^2(1+z^2)(E^2-\Delta ^2)}}{\displaystyle{
(E+(1+2z^2)\sqrt{E^2-\Delta ^2}))^2}}\\ 
T_0=\frac{\displaystyle{
2E(1+z^2)(E+\sqrt{E^2-\Delta ^2})}}{\displaystyle{
(E+(1+2z^2)\sqrt{E^2-\Delta ^2}
))^2}} \\
T_a=\frac{\displaystyle{
2Ez^2(E-\sqrt{E^2-\Delta ^2})}}{\displaystyle{
(E+(1+2z^2)(\sqrt{E^2-\Delta ^2}))^2}} 
\end{array}
\end{equation}

The above treatment of a planar interface has been extended by a number of
authors, including [Riedel and Bagwell 1993]. In one dimension and at zero
temperature,
the boundary conductance
of equation (\ref{a21}) reduces to the BTK result

\begin{equation}
\label{5.11}
G=(2e^2/h)(1-R_0+R_a),
\end{equation}
which for $E<\Delta$
 becomes $G=(2e^2/h)2R_a$,
while for $E\gg \Delta $, $R_a\rightarrow 0$ and
one  recovers the Landauer expression (2), $G=(2e^2/h)(1-R_0)=(2e^2/h)T_0$.
It
is interesting to compare the sub-gap conductance
\begin{equation}
\label{5.12}
G_{NS}=(2e^2/h)\frac 2{(1+2z^2)^2} 
\end{equation}
with that in the normal state
\begin{equation}
\label{5.13}
G_N=(2e^2/h)T_N=(2e^2/h)\frac 1{1+z^2}.
\end{equation}
This allows one to express $G_{NS}$ in terms of the dimensionless
normal state conductance $T_N$:
\begin{equation}
\label{5.14}
G_{NS}=(2e^2/h)\frac{2T_N^2}{(2-T_N)^2},
\end{equation}
which is a consequence of the fact that the
superconducting electrode in this approximation only enters as a 
boundary condition.

For a disordered N-region in contact with
a superconductor the result (\ref{5.14})
can be extended to higher dimensions by introducing the eigenvalues
 $T_n$
of the transmission matrix $T=t_{pp}t^\dagger_{pp}$ of the N-region and
summing equation (\ref{5.14}) over all $n$ [Beenakker 1992], to yield
equation (\ref{5.19}) below.
For $z\neq 0$ , equation (\ref{5.11})
predicts that the low-energy conductance is suppressed by a
factor $(1+z^2)^{-2}$, as compared to the $z=0$ case.
Already for values of $
z\sim 3$ this implies an almost vanishing sub-gap conductance. At the gap
energy, the expression for $R_a$ achieves a peak  value of unity
 within Andreev
approximation, with $R_a$ decaying to zero for energies above the gap. 
This picture clearly resembles that of
 the tunneling approach, where the conductance is controlled by the
supeconducting density of states, but is at variance with
the experiment of
[Kastalskii et al. 1991]. It was soon recognized that
the  low-energy peak observed by Kastalskii et al
was due to Andreev scattering [van Wees, de Vries,
 Magn\'ee, and Klapwijk, 1992], but it was
not possible to explain it using the theory discussed above,
because  the interplay between  scattering due
to disorder in the normal region and Andreev scattering at the N-S interface
 is crucial.  The  first insight into the ZBA was based on
a  detailed description using quasi-classical Green function
methods [Zaitsev 1990,
Volkov and Klapwijk 1992, Volkov, Klapwijk and Zaitsev 1993].
Hekking and
Nazarov [ 1993,1994], obtained similar results using a
tunneling approach
and Beenakker and co-workers [Beenakker et al 1994; Marmorkos et al 1993]
confirmed these
results using multiple scattering methods.

To illustrate how these different
methods explain the ZBA, the following sub-section briefly summarizes
quasi-classical predictions for the zero-energy
conductance and in sub-section C, the multiple scattering
approach is discussed.
%%%%%%%%%%%%%%%%%%%%%%%%%%%%%%%%%%%%%%%%%%%%%%%%%%%%%%%%%%%%%%%%%%%%%%
\subsection{The quasiclassical approach: circuit theory.}

To apply the ''circuit rules'' of  quasi-classical theory[Nazarov 1994]
 to a N-I-S system, we  consider two resistances in series;
the first is associated with the normal disordered region (N),
while the second is associated with the tunnel junction (I). We assume that
 the normal region
progressively widens in a macroscopic electrode and that the 
superconductor
plays the role of the other electrode. In this circuit there are three
spectral vectors. The vector associated with the normal electrode
is ${\bf s}_N={\bf z}$
 and that associated with the superconducting electrode
is ${\bf s}_S={\bf x}$, both or which are
 fixed (cf. rule II of circuit theory). The third spectral vector is
 located  at
the node connecting the diffusive element and the tunnel junction and
lies in the $x-z$ plane
 ${\bf s}=\cos (\theta ) {\bf x} +\sin (\theta ) {\bf z}$. 
According to the circuit rules, one
 computes the conductance of this simple circuit in the usual way,
but the resistance of the tunnel junction is renormalized
(cf. rule I of circuit theory)
\begin{equation}
\label{5.15}
G_{NS}=\frac 1{1/G_D+1/(G_T\sin (\theta ))} 
\end{equation}
where $G_D$ and $G_T$ are the conductances of the normal diffusive region
and of the tunnel junction in the normal state, respectively. The angle
$\theta$ is determined from the conservation of the vector current 
(cf. rules III-IV of circuit theory) and
satisfies the equation

\begin{equation}
\label{5.16}
G_D\theta =\cos (\theta ) G_T.
\end{equation}

This equation has two simple limits: i) $G_D/G_T \gg 1$,
 $\theta \approx G_T/G_D$,
$G_{NS}\approx G_T^2/G_D$; ii) $G_D/G_T \ll 1$, $\theta \approx \pi /2$,
$G_{NS}\approx G_D +G_T$.
This shows that, by varying the parameter $G_T/G_D$, the conductance
 switches from
a quadratic to a linear dependence on the tunnel junction conductance.
 A two-particle
process behaves effectively as a one-particle process 
($G_{NS} \approx G_T$). The smallness
of the probability for a two-particle transmission is compensated by
 the relative
increase of disorder-induced scattering on the normal side.
A detailed comparion between eq.(\ref{5.15}) and a numerical 
S-matrix investigation can be found in 
Claughton, Raimondi and Lambert [1996].
%%%%%%%%%%%%%%%%%%%%%%%%%%%%%%%%%%%%%%%%%%%%%%%%%%%%%%%%%%%%%%%%%%%%%%
\subsection{The scattering approach: random matrix theory.}

To illustrate the multiple scattering approach to the ZBA,
we now show how equation (\ref{5.15}) may be derived
using random matrix theory [Beenakker et al 1994; for a more detailed introduction to random
matrix theory  applied to quantum transport see also 
Stone, Mello, Muttalib and Pichard [1991] and the
recent review by
Beenakker 1997].
The starting
point is the following scaling equation for the eigenvalue density
$\rho (\lambda , s )=<\sum_{n=1}^N \delta (\lambda -\lambda_n )>$
[P.A. Mello and J.-L. Pichard 1989],

\begin{equation}\label{5.17}
{{\partial}\over {\partial s}}\rho (\lambda , s)=
-{2\over N} {{\partial}\over {\partial \lambda}} \lambda (1+\lambda )
\rho (\lambda , s) {{\partial}\over {\partial \lambda}}
\int^{\infty}_{0}d\lambda ' \rho (\lambda ', s) \ln |\lambda -\lambda '|
\end{equation}

where $\lambda_n$ is related to the transmission eigenvalue $T_n$ of the n-th transverse
mode by

\begin{equation}\label{5.18}
\lambda_n = (1-T_n)/T_n
\end{equation}
and the average $<...>$ is over the appropriate ensemble.
Here  $s=L/l$, where $L$ is the size of the system and $l$ the mean
free path.  $N$ is the number of open channels in the normal region.
The above equation yields the ensemble averaged eigenvalue density
for a diffusive system and it is physically equivalent to the quasiclassical
theory outlined in the previous section.  Once
$\rho (\lambda , s)$ is known,
the conductance of the N-I-S system is given by
(cf. eq.(\ref{5.14}))

\begin{equation}\label{5.19}
G_{NS}= {{2e^2}\over h}\sum_{n=1}^N {{T_n^2}\over{(2-T_n)^2}}=
{{2e^2}\over h} 2\int^{\infty}_0 d\lambda \rho (\lambda , s)
{1\over {(1+2\lambda})^2}.
\end{equation}
With respect to the scaling variable $s$, the presence of a barrier at the
N-S interface, acts as an "initial condition", instead of a "boundary
condition". In fact, at the interface, where $s=0$,
the eigenvalue
density in a mode-independent approximation, is given by

\begin{equation}\label{5.20}
\rho (\lambda , 0)=N \delta (\lambda -(1-T)/T),
\end{equation}
where $T$ is the transmission coefficient of the barrier.
To solve the evolution equation (\ref{5.17})
 with the  initial condition (\ref{5.20})
one introduces the auxiliary function $F(z,s)$
\begin{equation}\label{5.21}
F(z,s)=\int^{\infty}_0 d\lambda ' {{\rho (\lambda ', s)}\over
{z-\lambda '}},
\end{equation}
which is analytic in the $z$-complex plane with a cut along the positive
real axis. In terms of $F$, the evolution equation (\ref{5.17}) becomes
\begin{equation}\label{5.22}
N{{\partial}\over {\partial s}}F+{{\partial}\over {\partial z}}z(1+z)F^2=0.
\end{equation}

Writing $z=\sinh^2 (\xi )$ and
\begin{equation}\label{5.23}
U(\xi , s) = {{\sinh (2\xi )}\over {2N}}F(z(\xi ),s)
\end{equation}
yields
\begin{equation}\label{5.24}
{{\partial}\over {\partial s}}U+U{{\partial}\over {\partial \xi}}U=0,
\end{equation}
which  is  Eulers equation of an ideal two-dimensional fluid.
The real and imaginary parts of $U$ are the cartesian components of the
velocity field $U=U_x +i U_y$. The solution can then be obtained directly
in terms of the initial condition $U_0 (\xi )\equiv U(\xi , s )$, as
\begin{equation}\label{5.25}
U(\xi , s)=U_0 (\xi -sU(\xi , s)).
\end{equation}

The initial condition (\ref{5.20}) 
for the eigenvalue density translates for $U$ in 

\begin{equation}\label{5.27}
U_0(\xi )={{\sinh (2\xi )}\over {2(\cosh^2 (\xi ) -T^{-1})}},
\end{equation}

which in the limit of small transparency barrier, i.e., $T\ll 1$
is approximated by (cf. eq.(\ref{4.57}))
\begin{equation}\label{5.28}
U_0 (\xi ) =-{T\over 2}\sinh (2\xi ).
\end{equation}

To evaluate the conductance, we note
that in terms of the function $U$,
 the equation (\ref{5.19}) becomes
\begin{equation}\label{5.26}
G_{NS}={{2e^2}\over h} N\lim_{\xi\rightarrow -i\pi/4} 
{{\partial U}\over {\partial \xi}}.
\end{equation}
By using (\ref{5.25}) together with (\ref{5.28}) and writing  the real and
immaginary part of $U$ at $\xi =-i\pi/4$, we obtain the pair of equations

\begin{equation}\label{5.29}
\begin{array}{cc}
U_x&=-{\displaystyle{T\over 2}} \sinh  (2sU_x ) \sin (2sU_y) \\
U_y&={\displaystyle{T\over 2}} \cosh (2sU_x ) \cos (2sU_y).\\
\end{array}
\end{equation}
The first equation requires $U_x =0$, and by defining $\theta =2sU_y$,
we rewrite the second equation
\begin{equation}\label{5.30}
\theta=sT\cos (\theta ).
\end{equation}

By recalling that $G_D =(2e^2/h)( l/L)$ and $G_T =(2e^2/h) T $ are the conductances of the disordered normal region and of the tunnel barrier, we recover the equation for the conservation of the vector current 
(\ref{5.16}) obtained within circuit theory.
The expression for the conductance (\ref{5.26}) may now be obtained by noticing that
\begin{equation}\label{5.31}
{{\partial U}\over {\partial \xi}}={1\over {s-{1\over
\displaystyle{{T\cosh (2\xi -2sU )}}}}}
\end{equation}
and setting $\xi =-i\pi/4$. Insertion of equation(\ref{5.31}) into
equation(\ref{5.26}),
leads to an expression for the conductance identical to
that given by circuit theory (cf. eq.(\ref{5.15})).

The above ideas have been further developed in a number of treatments of
Andreev scattering in chaotic and resonant stuctures, including
[Altland and Zirnbauer 1995, Berkovits 1995, Beenakker et al 1995,
Brouwer and Beenakker 1995(a,b), Bruun et al 1995,
Claughton et al 1995(b), 
Argman and Zee, 1996, Fraum et al 1996, Melsen et al 1996,
 Slevin et al 1996]

%\end{document}

%  REENTRANT PHENOMENA,
\section{Reentrant and long-range proximity effects.}
\subsection{Reentrant behaviour of the conductance.}

As well as  the ZBA discussed in the previous section,
a number of recent
experiments have revealed a non-monotonic behaviour of
the voltage and temperature dependence of the conductance. In this section,
we use the quasi-classical approach to highlight the origin of this
phenomenon. For simplicity, we consider the case of a normal
diffusive wire located in the region $0 < x <L$
 between a normal and a superconducting reservoir.
 As explained in section IV(C) the equation
for the Green's function $F^R$, which is conveniently written

\begin{equation}
\label{6.1}
D\partial _x^2u(x)+2i\epsilon \sinh (u(x))=0,
\end{equation}
must be solved with boundary conditions $u(0)=0$ and $
u(L)=i\pi /2$. The voltage dependent conductance is then obtained from
\begin{equation}
\label{6.2}
\frac{G(\epsilon )}{G_o}=\left\{ \frac 1L\int_0^Ldx\frac 1{\cosh
{}^2(Reu(x,\epsilon ))}\right\} ^{-1} 
\end{equation}

where $G_o$ is the conductance in the normal state.
Using the identity $\cosh {}^{-2}w=1-\tanh {}^2w$,
 equation (\ref{6.2}) becomes
\begin{equation}
\label{6.3}
\frac{G(\epsilon )}{G_0}=\left\{ 1-\frac 1L\int_0^Ldx\tanh
{}^2(Reu(x,\epsilon ))\right\} ^{-1}. 
\end{equation}
When $Reu(x,\epsilon )$ is
 a small quantity, one can
expand the denominator of equation (\ref{6.3}) to yield
\begin{equation}
\label{6.4}
\frac{\delta G(\epsilon )}{G_o}=\frac 1L\int_0^Ldx\tanh {}^2(Reu(x,\epsilon
)),
\end{equation}
which allows us to obtain
 approximate solutions in both the low-
and high-energy limits. In the limit $\epsilon \ll \epsilon _T$ where $
\epsilon _T=D/L^2$, one can treat the term linear in the energy as a
perturbation and write $u(x,\epsilon )=u_o(x)+\delta u(x,\epsilon )$. When
inserted into  equation (\ref{6.1}), this yields
\begin{equation}
\label{6.5}
u(x,\epsilon )=
-i\frac \pi 2\frac xL+\frac 8{\pi ^2}\frac{\epsilon }{
\epsilon _T}\left( \sin (\frac{\pi x}{2L})-\frac xL\right) . 
\end{equation}
In the opposite limit of large energies, $\epsilon \gg \epsilon _T$, one
notices that among the solutions of the original non-linear equation, there
exists a subset satisfying the following equation
\begin{equation}
\label{6.6}
\partial _xu=2k\sinh (\frac u2). 
\end{equation}
Solutions of   equation (\ref{6.6}) are also solutions of the original
 equation (\ref{6.1})
 \footnote{Although the converse
 is not generally true, because there may be solutions of the original
equation, which are not solutions of the above equation. This in general will
depend on the boundary conditions, as is evident from the fact that the
original equation is second order while the derived above equation is first
order.}. The solution of equation (\ref{6.6}) reads
\begin{equation}
\label{6.7}
\tanh (\frac{u(x,\epsilon )}4)=\tanh (\frac{u(L,\epsilon )}4)e^{k(x-L)} 
\end{equation}
where $k=\sqrt{-2i\epsilon /D\text{.}}$ At the normal reservoir, the value
of the solution cannot be imposed, though in the limit of large energies,
the r.h.s. of (\ref{6.7}) 
is exponentially small and the solution satisfies the condition
that the pairing function must vanish. We also note that in both the low- and
high-energy limits, the real part of $u(x,\epsilon )$ is a small quantity
and therefore the assumption made at the beginning is justified. With these
two
solutions we may finally write
\begin{equation}
\label{6.8}
\frac{\delta G(\epsilon )}{G_o}=A\left( \frac \epsilon {\epsilon _T}
\right)
^2,\epsilon \ll \epsilon _T 
\end{equation}
\begin{equation}
\label{6.9}
\frac{\delta G(\epsilon )}{G_o}=B\sqrt{\frac{\epsilon _T}\epsilon },
\epsilon
\gg \epsilon _T, 
\end{equation}
with $A$ and $B$ numerical constants.

The above analysis illustrates  the key ingredients required
to obtain a reentrant effect, which was
noted prior to the experiments
by [Artemenko, Volkov and Zaitsev 1979].
The effect originates from the presence of
a distribution function which
is spatially not in equilibrium and consequently
will be most relevant in situations where the voltage drop
is distributed along the wire, rather than dropping predominantly
across the tunnel junction.

The fact that the conductance variation appears to be a small quantity
both in the low energy and in the high energy limit, 
controlled respectively by
the small parameters $(\epsilon /\epsilon_{T})^2$ and
$\sqrt{\epsilon_{T} /\epsilon}$, is not accidental. It has been noticed
that the conductance correction resembles the fluctuation correction
in the theory of paraconductivity, the so-called Maki-Thompson term
[Volkov 1994(a)]. In the high energy limit, the proximity effect induces
 Cooper pairs in the normal metal, which act like a superconducting
fluctuation above the critical temperature. At low energy, the induced
pair wave function is not small, so at first sight
the  correction should be sizeable. However the "longitudinal"
component of the induced pair wave function gives a contribution
which is equal to the normal state conductance and is not temperature dependent,
while the "transverse" component, which is small at low energies, gives
the additional temperature dependent contribution.
On general grounds, one expects that the low energy
$\epsilon^2$ dependence should be universal, while the high energy behaviour
may differ from the inverse square root law, depending on the quality
of the interface and on the effective dimensionality of the mesoscopic
wire.

\subsection{Long-range proximity effects.}

We now discuss briefly the origin of the long-range
proximity effects,  observed in the experiments of
[Petrashov et al. 1993,1995,
Dimoulas et al. 1995, Courtois et al. 1996],
where the conductance of a mesoscopic
loop between two superconducting electrodes was measured.
In such structures, the total current is the sum of two terms.
The first is a Josephson current between the two superconducting
electrodes and  decays exponentially as $L/L_T$ in agreement with
known theory of superconducting weak links [Likharev 1979].
The second  was shown to be due to a phase-periodic
conductance with amplitude decaying as the inverse of the temperature. 
A detailed theory of the effect has been provided by [
Volkov, Allsopp and Lambert 1996 and Volkov and
Takayanagi 1996(a)] using both quasi-classical and multiple scattering
methods. The different
behaviour of the two contributions to the current occurs because
the Josephson current has a thermodynamic
origin and is obtained by integrating over
a range of energies of the order of $k_BT$, while
the second contribution  giving rise to the
reentrant behaviour is kinetic in nature. In the structure considered
in the experiments,  a non-vanishing
value of the pair wave function $F(x,\epsilon )$ is
induced in the normal region. Away from the
superconductor the pair wave function decays exponentially as

\begin{equation}
F (x, \epsilon )\approx exp(-kx )
\label{8.1}
\end{equation}

where $k=\sqrt{\epsilon /D}$.  The contribution to the Josephson
critical current comes from an integral over the energy involving
the product of anomalous Green's function $F^R F^R$ and $F^A F^A$.
Since this product contains Green's functions
with poles on the same side of the real axis, one can deform the
contour of integration and transform the integral over the energy 
 to a sum over
Matsubara frequencies $\epsilon_n =\pi T (2n+1)$. As a result, the
main contribution comes from energies of the order of the temperature.
At a distance L, the Josephson current decays exponentially as
$L/L_T$ where $L_T =\sqrt{D/T}$. By introducing the Thouless energy
$\epsilon_T =D/L^2$, the condition for a vanishing Josephson 
current becomes $\epsilon_T \ll T$. In addition to the above,
the proximity effect gives rise to an additional contribution
to the normal state conductance, which involves products of the type
$F^R F^A$.  As a consequence, the integral over the energy
cannot be transformed to a sum over Matsubara frequencies. (This
 expresses the fact that the main contribution to the integral
comes from energies of the order of the Thouless energy.) As
a consequence, if the
condition $\epsilon_T \ll T$ is satisfied, then the pair wave function
spreads over the entire sample.

%INTERFEROMETERS.
\section{Andreev interferometers}

When a quasi-particle Andreev reflects from a normal-superconducting
interface, the phase of the outgoing excitation is shifted by the phase
of the superconducting order parameter. Consequently if a phase-coherent
normal conductor is in contact with two superconductors with order 
parameter phases $\phi_1$ and $\phi_2$, transport properties will
be oscillatory functions of the phase difference $\phi =\phi_1 -\phi_2$.
Following a number of
theoretical proposals [Spivak and Khemel'nitskii 1982, Altshuler and
Spivak 1987, Nakano and Takayanagi 1991, Takagi 1992,
Lambert 1993, Hui and Lambert 1993(a), Hekking and Nazarov 1993],
several  experimental realizations of Andreev interferometers
have been reported
 [de Vegvar et al. 1994, Pothier et al. 1994,
 van Wees et al 1994,
Dimoulas et al. 1995, Petrashov et al. 1995, Courtois et al 1996].
In addition to interferometers formed when a normal metal makes contact
with two superconductors,  the conductance
of a single exended N-S structure
is predicted to be an oscillatory function of the
phase gradient across the interface [Cook et al 1995], although
no experiments on such phase-gradiometers have been reported to-date.

These papers address a number of issues
including the fundamental periodicity
of the conductance oscillation, the nature of the zero-phase extremum,
the magnitude of the
oscillation amplitude and the role of disorder, geometry and dimensionality.
In the main, experiments have probed either diffusive or ballistic structures,
which in-turn can be divided into metallic samples with a conductance
$G$ much greater than $e^2/h$ and semiconducting structures with $G$
less than or of order $e^2/h$. For the former, the ensemble averaged
conductance $<G>$ is the relevant quantity and mesoscopic fluctuations
are unimportant.
These have confirmed the $2\pi$ periodicity for $<G>$ predicted by
[Lambert 1993, Hui and Lambert 1993] and by all subsequent theories,
 [Hekking and Nazarov 1993, 1994,
Zaitsev 1994, Takane 1994, Allsopp et al  1996, Kadigrobov et al 1996,
Nazarov and Stoof 1996,
Volkov and Zaitsev 1996, Zagoskin et al 1996, Leadbeater and Lambert 1997].
For the latter, mesoscopic  fluctuations
can dominate and the non-averaged conductance $G$ is of interest,
which depends in detail on the geometry and impurity realisation of
a particular sample. These confirm the $2\pi$ perodicity for the
non-averaged conductance predicted initially for diffusive systems
by [Altshuler and Spivak 1987] and for one-dimensional clean systems
by [Nakano and Takayanagi 1991 and Takagi 1992].

In this section we briefly review the
various theoretical approaches and  comment on their
applicability to the experiments.

\subsection{Quasi-classical theory: the dirty limit.}

We start by using quasi-classical theory to compute the ensemble
averaged conductance
when the mesoscopic normal conductor is in the diffusive
transport regime. As noted
in previous sections, the influence of superconductivity
on the transport properties of a mesoscopic conductor may change
in a qualitative way, depending on the quality of the
normal-superconductor interfaces. 
Consider first the zero temperature conductance of
an interferometer
comprising a tunnel junction (labelled as 1) connected by 
diffusive 1-D wires to a fork.
Each of the two arms of the fork is a diffusive wire, connected via 
tunnel junctions (labelled as 2) to infinitely long superconductors. The conductance
of the diffusive wires is assumed to be much greater than that of the
tunnel junctions. In this structure the superconductors play the role
of elctrode and the main voltage drop occurs at the tunnel barriers.
The conductance can be analysed
using the circuit theory outlined in section IV
[Nazarov 1994, Zaitsev 1994], which yields

\begin{equation}
\label{7.1}
G = \frac
{4 \, G_1^2 \, G_2^2 \, \cos^2 (\phi/2) }
{\left\{ G_1^2 + 4 \, G_2^2 \, \cos^2 (\phi/2) \right\}^{3/2}},
\end{equation}

where $G_1$ is the conductance of the tunnel junction (1), $G_2$ is the
conductance of the tunnel junction (2) and $\phi$ is the phase difference
between the two superconductors.
In the limit $G_1 \gg G_2$, this simplifies to

\begin{equation}
\label{7.2}
G = 4 \, \frac{G_2^2}{G_1} \cos^2 (\phi/2),
\end{equation}

whereas if $G_2 \gg G_1$,

\begin{equation}
\label{7.3}
G = \frac{1}{2}\,\frac{G_1^2}{G_2} \,\frac{1}{|\cos^2 (\phi/2)|}.
\end{equation}

 One can see that for $G_1 = G_2$ there  is a zero-phase minimum, 
as there is for $G_1 \ll G_2$.  However when $G_1 \gg G_2$,
there arises a zero-phase maximum. In all cases the conductance
vanishes when the phase difference between the superconductors is $\pi$. 
Furthermore, the ratio $G_1 / G_2$ controls the amplitude of the
conductance  oscillations which are greater when $G_1 \approx G_2$.

The results of quasi-classical theory can of course be compared
with the results of numerical multiple scattering theory.
If the parameters used
in the numerical simulations are chosen in such a way to satisfy the
assumptions underlying the quasi-classical treatment, then
a detailed analysis shows that
this is indeed the case [Claughton, Raimondi and Lambert 1996],
although certain features, such as the vanishing of the conductance
at $\phi =\pi$, appear to be an artifact of te one-dimensional
nature of the analysis leading to equation (155).
It is perhaps worth emphasizing that the exact numerical techniques
used in [Claughton, Raimondi and Lambert 1996]
are not confined to the diffusive regime
and an interesting result of  these simulations is that when the
resistance of a given structure is dominated by tunnel barriers, the
diffusive
nature of the wires is not relevant and similar results are
obtained for both ballistic or diffusive wires.
The geometry is clearly important, but
when dissipation occurs along the wire, the almost
one-dimensional approximation used in many quasi-classical calculations
works well.

Another structure illustrating the role of finite temperature
and voltage
is suggested by the experiment
 by Petrashov et al. [1995], who measured the conductance
of a mesoscopic wire, which makes contact with superconductors
at points between the normal electrodes. At zero temperature,
this structure is predicted to have only small amplitude of osicllation arising
from mesoscopic flcutuations, in contrast with the large amplitude
oscillations observed experimentally. In the experiment however,
a finite voltage drop is distributed along the wire and the temperature
 is non-zero. Consequently the
phase periodic conductance is dominated by an effect similar
to the reentrant behaviour discussed in the previous section.
The analysis of this structure has been carried out in detail by
[Nazarov and Stoof 1996, Volkov, Allsopp and
Lambert 1996, Stoof and Nazarov 1996], using both
quasi-classical and numerical multiple scattering methods.

\subsection{Description of the ballistic limit: a simple
two-channel model.}

Having discussed diffusive interferometers, we now examine the clean limit,
which was first described in one dimension by
[Nakano and Takayanagi 1991 and Takagi
1992].
An analytic theory of the
ballistic limit in higher dimensions has also been developed
[Allsopp et al 1996],
based on a  multiple scattering description of a clean
N-S interface.  In the absence of disorder,  when there
is no phase difference between the two superconductors, translational
 invariance in the direction parallel to the interface
 (and transverse to the current flow) allows one
 to reduce the two-dimensional system to the sum over many independent 
one-dimensional channels. When a phase difference between the 
superconductors is imposed, inter-channel coupling is introduced, 
but as shown below, this coupling typically involves only pairs of
channels in such a way that an accurate  description is obtained by
summing over independent {\bf pairs } of coupled channels.
This considerably simplifies  evaluation
of  the boundary conductance

\begin{equation}
G_{\rm B}(\phi)=2R_a=2{\rm Tr}\,\, r_ar_a^\dagger=
2\sum_{i,j=1}^N(R_a)_{ij}
\label{7.4},
\end{equation}
where $(R_a)_{ij}=\vert(r_a)_{ij}\vert^2$ is the Andreev reflection
probability from channel $j$ to channel $i$.
Indeed the Andreev reflection coefficient is of the form
$R_a=R_{\rm diag}+R_{\rm off-diag}$
where 
$R_{\rm diag}=\sum_{i=1}^N (R_a)_{ii}$ and
$R_{\rm off-diag}$ is the remaining contribution from
inter-channel scattering,
$R_{\rm off-diag}=\sum_{i\ne j=1}^N (R_a)_{ij}$. If channels only
couple in pairs, both the off-diagonal scattering and
the diagonal scattering will scale as the number
of channels.
Consider now a normal barrier in the N-region of a N-S interface.
Particles (holes) impinging on the normal scatterer are described by
a scattering matrix $s_{pp}$,
$(s_{hh})$, 
and those arriving at the N-S interface by a reflection matrix $\rho$,
where

$$
s_{pp}=\left(
\matrix{
r_{pp} &t'_{pp}\cr
t_{pp} &r'_{pp}\cr
}\right)
,~~~~
\rho=\left(
\matrix{
\rho_{pp} &\rho_{ph}\cr
\rho_{hp} &\rho_{hh}\cr
}\right).$$
The elements of $s$ and $\rho$ are themselves  matrices  describing
scattering between open
channels  of the external leads. For an ideal interface,
where Andreev's approximation is valid, $\rho_{pp}$ and
$\rho_{hh}$
can be neglected and
as a consequence, $\rho_{hp}$ and $\rho_{ph}$ are unitary
and 
one obtains a generalization of a formula due to Beennakker [1992]
$
r_a=t'_{hh} \rho_{hp} M_{pp}^{-1}t_{pp}
,$
with
$
M_{pp}=1-r'_{pp} \rho_{ph} r'_{hh} \rho_{hp}.$
In contrast with the analysis of Bennakker [1992], where $\rho_{hp}$ is
proportional to the unit matrix, the interference
effect of interest here is contained in the fact that $\rho_{hp}$
induces off-diagonal scattering.
Substituting $r_a$ into equation (\ref{7.4}) and taking advantage of 
particle-hole symmetry at $E=0$, yields

\begin{equation}
G=2 Tr\left(T Q^{-1} T (Q^{\dagger})^{-1}\right)    
\label{7.5}
\end{equation}
where
$
Q=\rho_{ph}^t+(r')_{pp}\rho_{ph}(r')_{pp}^{\dagger},$
with $T=t_{pp}t_{pp}^{\dagger}$ the transmission matrix of the
normal scattering region. This multiple scattering formula for the
boundary conductance is valid in the presence of an arbitrary number of
channels and in any dimension.

Equation (\ref{7.5}) is very general and makes no assumption
 about the nature of matrices $\rho_{ph}$ and $s_{pp}$.
In a  two-channel model,
 $\rho_{ph}$  is chosen to be an arbitrary two dimensional unitary
matrix and since
in the absence of disorder,  $t_{pp}$ and $r_{pp}$ are
 diagonal and interchannel coupling arises from $\rho_{ph}$ only.
 Substituting these
matrices into equation (\ref{7.5}), yields an expression for
$r_a$ involving a single phase $\theta$, whose value is a linear 
combination of 
  phase shifts due to normal reflection at the barrier, Andreev 
reflection at the N-S interface and the phase
accumulated by an excitation travelling from the barrier to
the interface.
After averaging over the rapidly varying phase $\theta$, the diagonal
and off-diagonal terms are determined and found to vary periodically with
$\phi$ with period $2\pi$ as predicted in one dimension by
[Nakano and Takayanagi 1991]
A key new feature of equation (\ref{7.5})
 is that the in the absence of the normal
barrier, unitarity ensures the the total Andreev reflection obtained by
summing these two contribution is independent of $\phi$ and therefore
the amplitude of conductance oscillation vanishes.
Thus to obtain a maximum amplitude of oscillation, the tunnel barrier must
be tuned to a non-zero value.

An interesting mechanism capable of producing large conductance oscillations
which scale
with the number of channels has been proposed by [Kadigrobov et al 1995],
based on the simultaneous resonance of Andreev levels corresponding to
different channels. To illustrate this, consider for example
the clean
S-N-S structure analyzed by [Bardeen and Johnson 1972 and Kulik 1970].
For a long-junction, the Andreev levels have energies
\begin{equation}
E_n={{\hbar v_{F,z}}\over{2L}}[-\Delta \phi \pm (2n+1)\pi],
\label{bardeen}\end{equation}
where
$v_{F,z}$ is the component of the Fermi velocity  perpendicular to the
N-S interfaces and $\Delta\phi$ the phase difference between the
superconductors. In general these energies do not coincide,
but when $\Delta\phi =(2m+1)\pi$ all levels pass through the Fermi energy $E=0$,
yielding large-scale oscillations in
transport properties. In the calculation of [Kadigrobov et al 1995], normal
scattering arises from the beam splitters which form an integral
part of the structure
considered and therefore no addition tunnel barriers are needed to
 break the unitarity condition of [Allsopp et al 1996].

%. SUPPRESSION OF CONDUCTANCE
%\documentstyle[prl,aps]{revtex}
%\documentstyle[manuscript,aps]{revtex}
%
%\begin{document}
%\draft

\section{Conductance suppression by superconductivity.}

When a superconducting island is added to a normal host, one na\"{\i}vely expects
that the electrical conductance of the composite material will increase.
In contrast detailed calculations based on the multiple scattering formula
[Hui and Lambert 1993(b), Claughton et al 1995]
 predicted that the change $\delta G$ in the two
probe conductance of a mesoscopic sample, due to the onset of superconductivity
can have arbitrary sign. Indeed 
for a sample with a high enough conductance, a simple theorem
[Hui and Lambert 1993] states that $\delta G$
 is guaranteed to be
 negative and furthermore the magnitude of $\delta G$ is predicted to scale
 with $G$.
This \lq\lq anomalous proximity effect"
occurs in the absence of tunnel barriers
and at first sight appears to conflict with the quasi-classical
description presented in section IV, which predicts that at zero temperature,
the conductance
of equation (\ref{a22}), which takes the form $G=(4e^2/h)R_a$
is unchanged by the onset of superconductivity.
In this section we briefly discuss the origin of this effect and
how it can be reconciled with quasi-classical theory.

For convenience we separately
discuss the clean, diffusive and almost Anderson localised limits.

\subsection{The clean limit.}

The Landauer formula (2) predicts that the conductance of a normal
ballistic wire
 is $G=(2e^2/h) N$, where $N$ is the number of open scattering channels
at the Fermi energy $E_F$.
Consequently in the absence of spin-splitting,
if $E_F$ is varied, $G$ exhibits a series of steps of height
$2e^2/h$ corresponding to the opening or closing of scattering channels.
The effect of inducing a uniform superconducting order parameter
of magnitude $\Delta_0$ in such a wire
is shown in figure 9 of [Claughton et al 1995a]. In the vicinity of
normal-state steps, the conductance is predicted to be
extremely sensitive to the onset
of superconductivity. With increasing $\Delta_0$, the steps are destroyed and
$G$ decreases.
This behaviour is attributed to the breakdown of Andreev's approximation, which
is most pronounced for
scattering channels whose
wavevector $k_x$ normal to the N-S interface tends
to zero.
Such channels occur precisely at a normal-state conductance step.

For a clean  N-S-N structure, provided the
superconductor is much longer than the superconducting coherence
length $\xi = k_F^{-1}E_F/\Delta_0$, there is negligible quasi-particle
transmission through the S region and therefore the total
resistance reduces to the sum of two boundary resistances. Since the
boundaries are identical, this yields for the total conductance defined in
equation (\ref{a25}),
$G=R_a$ and since the system consists of decoupled
channels, $R_a$ can be obtained by solving the Bogoliubov - de Gennes
 equation at a one dimensional N-S interface. By insisting that scattered
  wavefunctions and their first derivatives be continuous at
the boundary, one finds [Claughton et al 1995a]
\begin{equation}
R_a = \sum_{n=1}^N R_n,
\label{9.1}\end{equation}

where
\begin{equation}
R_n={{2}\over{1+[1+(\Delta_0/\mu_n)^2]^{1/2}}}
\label{9.2}\end{equation}

and $\mu_n= \hbar^2{(k^x_n)}^2/2m$ is  the longitudinal kinetic energy of
a quasi-particle incident along channel $n$.
Clearly those channels corresponding to low angle quasi-particles
with $\mu_n < \Delta_0$,
 possess a small Andreev reflection probability $R_n$
  and since there is no transmission,
the corresponding normal reflection probability $(1-R_n)$ approaches unity.
Finally one obtains for  conductance change $\delta G$ due to the
switching on of a uniform order parameter in such a clean system,

\begin{equation}
\delta G= \sum_{n=1}^N [R_n-1]=
\sum_{n=1}^N {{1-[1+(\Delta_0/\mu_n)^2]^{1/2}}\over
{1+[1+(\Delta_0/\mu_n)^2]^{1/2}}}.
\label{9.3}\end{equation}

It is perhaps worth noting that although the suppression of Andreev
reflection for low-angle quasi-particles is quite general, the destruction
or otherwise of conductance steps in a quantum point contact, depends
on the geometry of the contact. If quasi-particles are adiabatically
accelerated before encountering the N-S interface, then Andreev reflection
need not be suppressed and conductance steps can survive. The criterion
for the survival of the $n$th step is clearly
\begin{equation}
\Delta_0/\mu_n << 1.
\label{9.4}\end{equation}

\subsection{The localised limit and resonant transport.}

The  results of the previous subsection
reveal that $\delta G$ is sensitive to
fine scale structure in the normal state conductance $G_N$. For clean systems this structure
takes the form of well-known conductance steps. For strongly
disordered systems, it is known that $G_N$ can exhibit sharp resonances
and  therefore the behaviour of $G$ with increasing $\Delta_0$
is sensitive to such features.
This behaviour is illustrated in figure 10 of [Claughton et al 1995],
which shows that negative changes in the conductance with increasing
$\Delta_0$ are associated with resonances in the normal-state transport.

In view of tunnelling theory, one might regard the suppression of conductance
in an almost insulating N-S structure as unsurprising, in which case the
increase in the conductance due to the onset of superconductivity
would be regarded as anomalous. A detailed description of such resonant
transport in N-S and N-S-N  structures and in resonant interferometers
is provided by [Claughton and Lambert 1995b], which follows
an earlier zero-voltage theory [Beenakker 1992] and a one-dimensional
description of a delta-like potential well [Khlus et al 1994].

\subsection{The diffusive limit.}

The above results show that the conductance of a clean or very dirty
conductor  can either increase of decrease when superconductivity is induced,
depending on the microscopic impurity configuration and on the
geometry.  This qualitative behaviour has been observed in experiments
by Petrashov and Antonov [1991] and Petrashov et al [1993(b)], which
also exhibit the weak magnetic field behaviour shown in fig 12 of
[Claughton et al 1995a]. Nevertheless a quantitative
 comparison between these calculations and  experiments is not possible,
 because of the differing geometries and  multiprobe measurements.

A first attempt at a quantitative comparison has been made recently by
Wilhelm, Zaikin and Courtois [1997] who noted that
although the one-dimensional approximation
 used in sec. V and VI to
study transport properties of N-S structures  successfully
explains a range of experiments,
the experiment by [Petrashov et al 1991]
requires a multiprobe description.
The key observation is that the
kinetic properties of a two-dimensional metallic film, in contact with a
superconductor, differ substantially from those of a quasi one-dimensional
wire, because of an inhomogenous distribution of the currents in the
sample.
Although a detailed explanation of the effect can be obtained by solving the
equations for the quasi-classical Green's function in a two-dimensional
structure, it is possible to gain considerable insight by adopting
an effective circuit model, which captures the main features. For example
consider four wires arranged in a square. The top two
 corners of the square (labelled A and B)
 are the voltage  probes, while the lower two corners
 (labelled C and D) are the current probes. Let the left-most vertical wire
 between corners A and C
be placed in contact with a superconductor and labelled 1,
the right-most vertical wire between corners B and D  labelled 2 and
the upper and lower wires labelled 3 and 4 respectively.
Then the four-probe
conductance reads

\begin{equation}
\label{fourprobes}
G_{four}=G_3 G_4(G_1^{-1}+G_2^{-1}+G_3^{-1}+G_4^{-1}),
\end{equation}

where  $G_i$ is the conductance of wire $i$.
The analysis of sections IV, V and
VI shows that the conductance of a wire in contact with a superconductor
is changed in several ways. 1) The local conductivity is {\sl increased}
by the presence of a superconductor, but this increase decays
over a distance from the superconductor of the order of the coherence length
of the normal metal $D/max\lbrace T,E\rbrace$. 2) The conductance has a non
monotonic behaviour as function of the energy and temperature.
To a first approximation one may consider that when
$T\gg D/L^2$,  only the wire attached to the superconductor will
be appreciably affected by the presence of the superconductor. 
In such a situation the conductances  $G_2$,  $G_3$ and  $G_4$ are the
same as in the normal state, while  $G_1$ is {\sl increased}.  In this case,
according to equation (\ref{fourprobes}), the four-probe conductance is
{\sl decreased} by superconductivity. On the other hand, at $T\ll D/L^2$,
all conductances increase and so does the four-probes conductance. As
[Wilhelm et al 1997] have shown this leads to a
 much richer structure in the temperature and voltage dependence of a 
two-dimensional film, compared with that of a one-dimensional wire
and to qualitative agreement with experiment.

A key prediction of [Wilhelm et al 1997] is that at zero bias
and zero temperature this effect should vanish and
the conductance should be unchanged by the onset of superconductivity.
In contrast [Seviour et al 1997] identify an additional large-scale effect
due to weak localisation, which is not
contained in standard quasi-classical theory
and which will survive at
zero-temperature and zero-bias.

%[ As for the mutliprobe section, I believe we should simply comment on
%these in the conclusion]
%\end{document}

%\section{Conclusions.}
\section{Conclusion.}

We conclude this review by indicating a few lines of future research.
The results of the first experiments in hybrid structures have been 
successfully explained in terms of almost one-dimensional models.
This has been possible because the physical phenomena responsible for
the transport properties of a given structure mainly occurred at the
N-S interface, so that the actual geometrical arrangement of current and
voltage probes was not important. 
This is clearly the case in the presence of low transmitting interfaces,
which tend to dominate the overall transport properties.
The improvement of the quality of N-S interfaces betweens separate parts of 
the hybrid structure and the contact leads makes necessary to take into
account the detailed geometrical arrangement of a given measurement setup
[Allsopp, Hui, Lambert and Robinson 1994].
Under this respect, the last twelve months have witnessed a considerable
experimental effort at paying particular attention to the geometry of a 
given measurement.  As we have seen in the previous section taking into
account the actual disposition of the current and voltage probes has been
crucial to describe the effect of the conductance suppression, even though
the structure itself has a one-dimensional geometry 
[Wilheml, Zaikin and Courtois 1997].  The recent experiment by 
[Hartog et al 1996] has beautifully shown the emergence of new physical
effects when the geometry of the experimental setup is considered carefully.

A second important semplifying assumption in examining the experiments
in hybrid structures consisted in neglecting the effect of electron-electron
Coulomb interaction. This is a reasonable assumption in many metallic and 
semiconducting wires used in the experiments. In these cases, a mean-field
treatment of electron-electron interaction is sufficient and can be easily
incorporated in the effective electronic parameters. 
In the last decade there has been however  a continuous progress in downsizing
fabrication of electronic systems,  so that it is now possible to measure
transport through almost zero-dimensional structures, or quantum dots.
In these systems, the effect of Coulomb interaction is no longer negligeable,
and various new effects arise.  Besides quantum dots, quantum wires also
have been a subject of noticeable attention. 
It is well known that in one-dimensional systems electron-electron interaction
drastically modifies the low energy properties and lead to the concept of
Luttinger liquid. Quantum wires are believed to be an experimental realization
of the Luttinger liquid state.  As a consequence in the last few years 
a considerable theoretical literature has accumulated with the intent of studing phase
coherent transport in interacting systems.  
It is almost a logical consequence to foresee in the forthcoming years the
emergence of a new field of research aimed at combining the fields of 
mesoscopic superconductivity and strongly interacting systems. Indees already
quite a few papers have started to appear 
[Fisher 1994,
Fazio, Hekking and Odintsov 1995, Maslov, Stone, Goldbart and Loss 1996,
Fazio and Raimondi 1997, Takane and Koyama 1997] and the time is mature for new
exciting developments.

\acknowledgements
Support from the EU TMR programme (Contract no. FMRX-CT 960042),
from the UK EPSRC and from the UK DERA is acknowledged.
%\end{document}

\begin{figure}
\caption{ Various generic experimental arrangements used in measuring dc transport
in superconducting hybrids.  The grey area indicates a  phase-coherent
normal region, whereas superconductive parts are marked by a S. The widening
open parts at the ends represent the reservoirs.}
\label{fig1}
\end{figure}

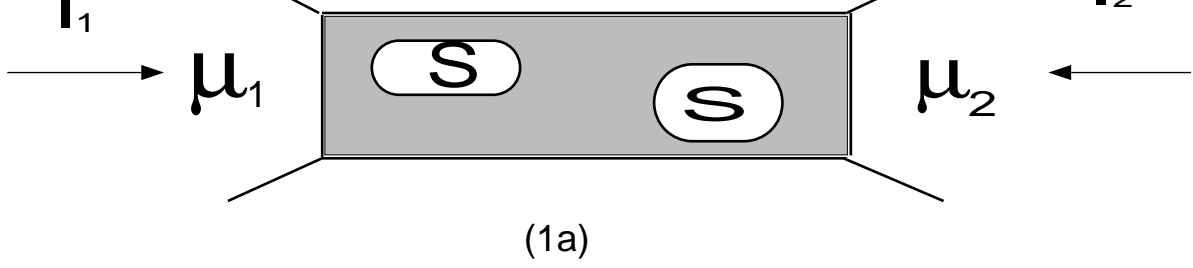
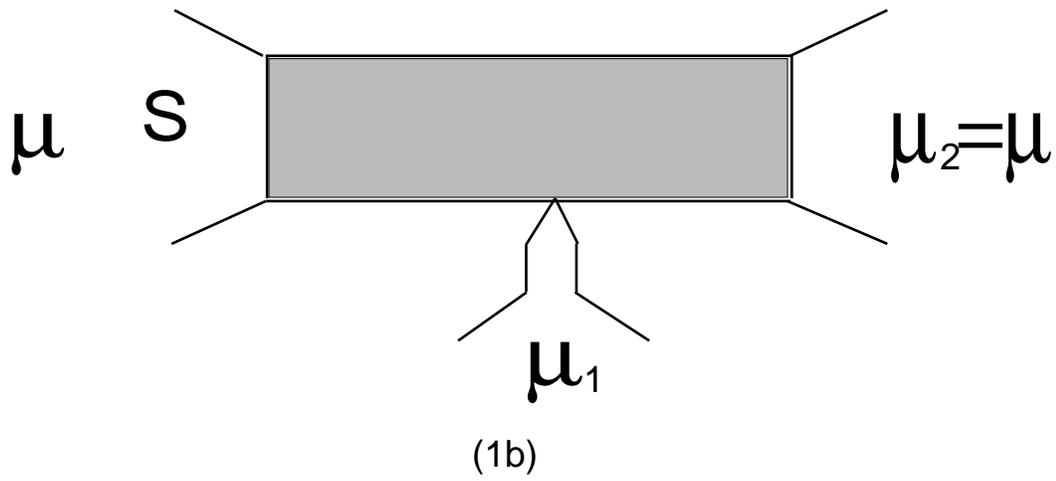
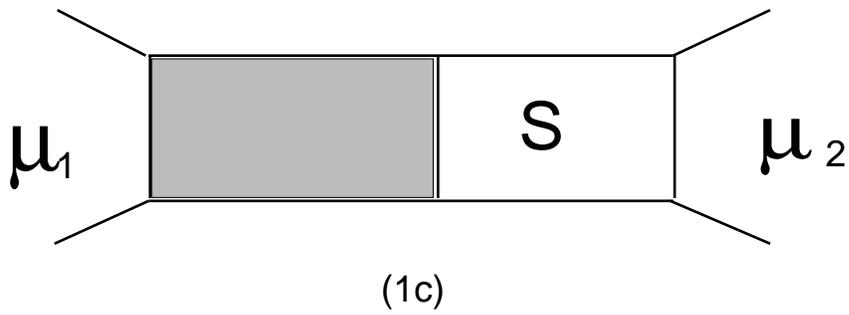
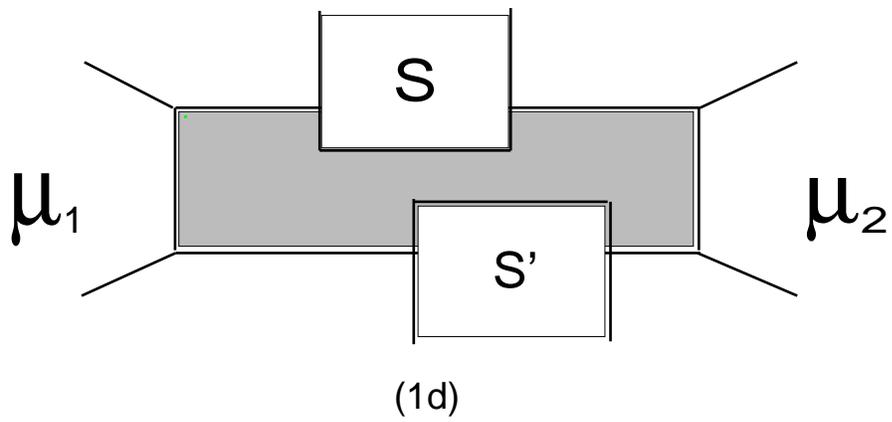
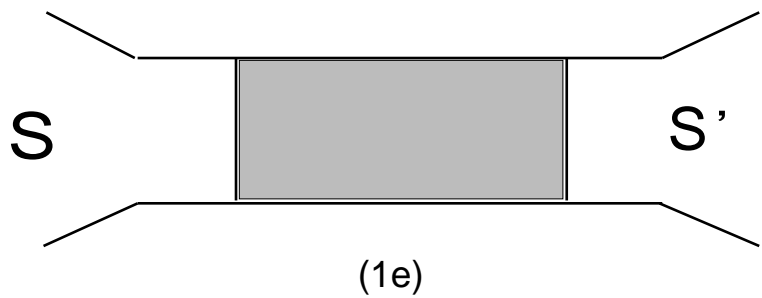
\end{document}